\documentclass[preprint,authoryear,12pt]{elsarticle}

\usepackage{amssymb}
\usepackage{amsmath}
\usepackage{graphicx}

\newcommand{\vx}{{\vec x}}
\newcommand{\cS}{{\cal S}}
\newcommand{\tL}{\widehat{\Lambda}}
\newcommand{\lve}{{\left(\frac{\varepsilon}{L}\right)}}
\newcommand{\btheta}{{\mathbf \Theta}}

\begin{document}

\begin{frontmatter}



\title{Multifractal point processes and the spatial distribution of wildfires in French Mediterranean regions}


\author[]{R. Ba\"{\i}le}
\ead{baile@univ-corse.fr}
\author[]{J.F. Muzy\corref{cor1}}
\ead{muzy@univ-corse.fr}
\author[]{X. Silvani}
\ead{silvani@univ-corse.fr}

\address{\footnotesize \em Laboratoire Sciences Pour l'Environnement \\

        \footnotesize \em CNRS UMR 6134 - University of Corsica \\

        \footnotesize \em Campus Grimaldi, 20250 CORTE (France)}

\cortext[cor1]{Corresponding author}


\begin{abstract}
We introduce a simple and wide class of multifractal spatial point patterns as Cox processes which intensity is multifractal, i.e., the class of Poisson processes with a stochastic intensity corresponding to a random multifractal measure. 
We then propose a maximum likelihood approach by means of a standard Expectation-Maximization procedure in order to estimate the distribution of these intensities at all scales. This provides, as validated on various numerical examples, a simple framework to estimate the scaling laws and therefore the multifractal properties for this class of spatial point processes. The wildfire distribution gathered in the Prom\'eth\'ee French Mediterranean wildfire database is investigated within this approach that notably allows us to compute the statistical moments associated with the spatial distribution of annual likelihood of fire event occurence. We show that for each order $q$, these moments display a well defined scaling behavior with a non-linear  spectrum of scaling exponents $\zeta_q$.  
From our study, it thus appears that the spatial distribution of the widlfire ignition annual risk 
can be described by a non-trivial, multifractal singularity spectrum and that this risk
cannot be reduced to providing a number of events per $km^2$.
Our analysis is confirmed by a direct spatial correlation
estimation of the intensity logarithms whose the peculiar slowly decreasing shape corresponds to the
hallmark of multifractal cascades. The multifractal features appear to be constant over time and similar over the three regions that are studied.
\end{abstract}





\end{frontmatter}



\section*{Highlights}
\begin{itemize}
  \item A multifractal spatial point process is defined as a spatial Cox process with a multifractal random intensity measure.
  
  \item The scaling and multifractal properties of this measure can be empirically estimated from samples by a maximum likelihood approach relying on the EM method. 
  
  \item The annual wildfire ignition spatial distribution over three south French Mediterranean regions can be represented within this framework and displays strong multifractal scaling properties that appear to be similar across these regions. 
  
  \item Our study suggests that the clustering features of the wildfire distributions do not result from peculiar correlations in the event occurrence likelihood but reflect the spatial structure of the intensity.
\end{itemize}

\newpage
\section{Introduction}
All over the world, each year, wildfires are responsible of high hazard and related damages with strong impact
on economic activity, biodiversity, decrease in forest, soil degradation and greenhouse effects.
Measuring and forecasting wildland fire risk is therefore of prime importance for safety management services.
This risk has mainly two components, namely the occurrence probability
and the hazard that accounts for the severity of the considered event through e.g., the burnt surface and its expected damages. In this paper, we focus on the risk related to the ignition probability which
has been at the heart of a wide number of studies over the past decade (see e.g., \cite{wf_review2012,wf_review2013,wf_review2015} and references therein). 
Many of these studies relie on statistical inference of spatial or spatio-temporal Point process models (see for example \cite{diggle_book,Gonzalez2016}) and mainly intended to capture the temporal and spatial dependences of fire occurrences
in order to evaluate or forecast the fire hazard and produce reliable operational information for help-to-decision making in prevention strategies (see, e.g., \cite{Genton2006,Hering2009,Turner2009,moller2010,xu2011,Ager2014,Serra2014,RODRIGUES201452,zhang2016,Opitz2017}).

Our purpose in this paper is to describe the spatial distribution of ignition risk at a coarser level and mainly to study its main statistical features through its spatial scaling properties. As many natural hazards, forest fires involve non-linear physical processes over a wide range of scales, yielding the scientific community to use tools and concepts from the physics of complex systems to study them. Within this framework, universality, scaling and self-similarity have proven to be fruitful concepts to account for many quantitative aspects of wildfire properties and to design pertinent phenomenological models. For instance, one of the main observation extensively studied during past years is the power-law behavior of the distribution of burned areas (\cite{Malamud09181998}) that has been considered within the theory of self-organized criticality (SOC) (\cite{Bak87}) or according to the mechanism of Highly Optmimized Toterance (HOT) (\cite{Hot00}). According to the first scenario, natural fires are examples of self-organized critical systems (\cite{Ricotta199973,Ricotta2001307,Turcotte2004580}) (i.e. a dynamical system ``naturally" behaving near a critical state) while the HOT theory assumes that the system is somehow ``optimized'' by a natural selection process in some state that makes it vulnerable to unusual conditions. Few studies suggest that scaling laws can also be observed in the dynamical aspects of wildfire occurrence (\cite{Corral08,Telesca2010}) or in their spatial distribution.
For instance, in \cite{TELESCA20071326,TUIA20083271}, the authors study the fractal nature of fire distribution patterns in central Italy by means of the correlation integral, box counting and sandbox methods
and found that ignitions are spread over a set of fractal dimension $D_c \simeq 1.5$ with a significant dependence on the burnt area. More recently, in \cite{KANEVSKI2017400}, the authors measure both global and ``local'' fractal dimension of
forest fire spatial distribution in Portugal and observed a non-homogeneity of the local scaling properties. If these studies suggest that wildfire ignitions appear to be heterogeneously distributed on a fractal set and propose various methods to estimate its dimension, the problem of the fractal or evenutually multifractal nature of wildfire spatial distribution remains overlooked to a large extent notably as far as mathematical and statistical issues are concerned.

In the present work, we ambition to bring a contribution to this field by studying scaling properties of wildfire 
ignition events through a simple yet general class of models for multifractal spatial point patterns. Our approach consists in considering the class of Cox processes whose intensity measure is a stationary random multifractal measure. Multifractal processes were introduced in the context of fully developed turbulence (see \cite{Fri95}) in order to describe objects that display non trivial scaling properties. They have been involved in many situations in both fundamental and applied disciplines. Notice that behind the notion of multifractal scaling there is often the picture of random multiplicative cascades according to which fluctuations are constructed by multiplication of successive fields of finer and finer characteristic scales. The paradigm that corresponds to such a picture is represented by so-called continuous, log-infinitely divisible cascades that display exact stochastic scale invariance properties (see for example \cite{BaMan02,mrw2,MuzyBaile16}). 
Spatio-temporal Cox processes have been considered in a many former works devoted to the modelling of forest fire distribution (\cite{moller2010,Serra2012,Serra2014,Pereira2013,Opitz2017}) in order to estimate and predict precise risk maps. Since our ambition is not to estimate 
any specific risk level at some precise location but to focus exclusively on the statistical laws at different scales,
we devise a simple maximum likelihood estimation procedure of the intensity distribution at all scales from which the moment scaling properties can be computed. This leads us to estimate not only the fractal dimension of the support of the event locations but also the singularity spectrum, a statistical measure introduced by \cite{FriPar85} in order to describe the fluctuations of local scaling properties. The relevance of this framework to study the wildfire spatial clustering properties is illustrated using the Prom\'{e}th\'{e}e database which contains the fire occurrences in three large regions of Southern France since 1973. 

The paper is organized as follows: in section \ref{sec:mpp}, we define
in which sense we consider that a point process is multifractal and introduce the class of 
multifractal Cox processes. We then describe a maximum likelihood method to estimate their multifractal scaling exponents. Our purpose is illustrated using simple examples of mono- and multi-fractal intensities spread over a Cantor set in 1D and 2D. The application of this approach to wildfire data is provided in section \ref{sec:multifracapp} where we consider the spatial variations of the ignition annual rate in the "Prom\'eth\'ee" database for 3 regions in the French Mediterranean area. After a brief description of the database and a discussion about the spatial clustering properties of annual fire event distribution, we model its empirical probability laws at different scales and study its multifractal
scaling properties that are compared for the 3 available regions.
In the concluding section, we comment our findings in terms of fire hazard and provide some prospects for future research. Some technical remarks about possible bias in fractal dimension and multifractal exponent estimations together with the analysis 
of clustering properties are provided in Appendices.

\section{Multifractal Cox processes and estimation of their scaling properties}
\label{sec:mpp}

\subsection{Multifractal spatial Cox processes}
Since a finite collection of sets of dimension $D$ has also 
a dimension $D$ and a single point has a dimension $D=0$,
a realization of a space-time point process over a bounded interval corresponds to set a dimension $D=0$. In that respect, when one refers to the fractal or multifractal nature of a point pattern there is a need to precisely define what one exactly refers to. In \cite{vj99}, the author shows that the (multi-) fractality of a space-time point process can be either considered from the point of view of the scaling properties of the associated spatial intensity process or from its clustering properties. In
the first case, one assumes that the process is ergodic in time and observed over a sufficiently long period so that one can estimate the spatial fluctuations of the intensity measure (also referred to as the expectation measure).
In the second situation, the process is assumed to be homogeneous in space but events occur in a
strongly correlated way, meaning that the so-called Palm distribution is slowly decreasing, e.g., behaves as a power-law.
This latter approach is the one that was notably taken in \cite{ogata91}. On a general ground, one can consider that the two previous features occur simultaneously and that the observed scaling properties is the intricate result of both intensity spatial inhomogeneities and event occurrence correlations.

In this paper, we will mainly consider the first scenario proposed in \cite{vj99} that basically consists in
neglecting possible correlations between event occurrence and focusing exclusively
on the peculiar spatial fluctuations of the expectation measure that encodes all the non-trivial
scaling features. As illustrated in \ref{clusterProp} (see also e.g. \cite{Hering2009}), it appears 
that spatial inhomogeneities in the occurrence rate (intensity) rather than spatial correlations and departures from 'Complete Spatial Randomness' can explain the observed clustering of annual forest fire events. In that respect, it is natural to consider the class of spatial Cox processes that are doubly stochastic Poisson processes with an intensity that is itself a stochastic process (see \cite{SPP_book,diggle_book} for the precise definition of a Cox process). Let us mention that Cox processes have been considered in the context of fire distribution 
modelling by various authors. They involve parametric spatio-temporal versions of 
the process (like Log-Gaussian with Mat\'ern function based kernel in \cite{Serra2012,Serra2014,Pereira2013,Opitz2017,Opitz2020} or Shot noise version in \cite{moller2010}) fitted using efficient statistical inference methods (\cite{diggle_book,Gonzalez2016}).
The objective of these studies is mainly to provide wildfire risk maps and fire risk prediction
by accounting for exogenous covariates like land use or climatic variables.

Our purpose is different since we do not consider such detailed spatial and temporal resolutions but rather focus on the scaling properties that characterize the spatial fluctuations of wildfire annual intensity considered as an homogeneous stationary process. Generally speaking,
we will consider a point process $dN(\vx)$ in $\mathbb{R}^d$ ($d=1,2,\ldots,$ being the dimension of the embedding Euclidean space) to be a spatial Cox process with an 
intensity that is provided by an homogeneous random multifractal measure $\Lambda$.
In particular, we suppose that, for any spatial domain ${\cal B} \subset \mathbb{R}^d$:
\begin{equation}
\label{eq:PoissonLaw}
  Prob \left\{ N({\cal B}) = n \right\}  = e^{-\Lambda({\cal B})} \frac{\Lambda({\cal B})^n} {n !}
\end{equation}
with $n \in \mathbb{N}$ and
\begin{equation}
  \Lambda({\cal B}) = \int_{\cal B} d \Lambda(\vx) \; .
\end{equation}
Let $B_\varepsilon(\vx)$ be a ball (or a square) of size $\varepsilon$ centered at position $\vx$.
If $d \Lambda(\vx) = \rho(\vx) d^dx$ (where $d^dx$ stands for the Lebesgue measure in $\mathbb{R}^d$), i.e., $d \Lambda$ is described by a density function $\rho(\vx)$, one has:
\begin{equation}
\label{eq:defmu}
   \rho(\vx)  =  \lim_{\varepsilon \to 0} \frac{\Lambda(B_\varepsilon(\vx))}{\varepsilon^d} = \lim_{\varepsilon \to 0} \frac{Prob \left\{N(B_\varepsilon(\vx)) = 1 \right\}}{\varepsilon^d}
\end{equation}
However, if the measure $d\Lambda(\vx)$ is singular as respect to the Lebesgue measure, the behavior $\Lambda(B_\varepsilon(\vx)) \sim \varepsilon^d$ does not hold and a density $\rho(\vx)$ cannot be defined. In that case, one has generically:
\begin{equation}
 \Lambda(B_\varepsilon(\vx))  \operatornamewithlimits{\sim}_{\varepsilon \to 0^+} \varepsilon^{\alpha(\vx)}
\end{equation}
where $\alpha(\vx) \in \mathbb{R}^{+\ast}$ is the local singularity exponent of $\Lambda$ at
position $\vx$. This is precisely
the situation that we want to take in account in this paper which occurs when
$\Lambda$ is a multifractal measure like a random multiplicative cascade. In this case, 
the local regularity is not the Euclidian dimension
($ \alpha = d$) and strongly varies pointwise.
The multifractal properties of $\Lambda$ can be described through the so-called {\em multifractal formalism} introduced in the context of turbulence (\cite{FriPar85,MS91}) and widely used in many areas ranging from chaotic dynamical systems to econophysics.
This formalism allows one to link the scaling properties of the measure with the statistical distribution of its singularity exponents $\alpha(\vx)$.
More precisely, one defines, for $q \in \mathbb{R}$, the {\em partition function}
\begin{equation}
\label{eq:def_zq}
  Z(q,\varepsilon) = \sum_{ i,B_\varepsilon(\vx_i) \in {\cal P}_\varepsilon} \Lambda(B_\varepsilon(\vx_i))^q
\end{equation}
where ${\cal P}_\varepsilon$ is a partition of $\cS$,
the observed support of $\Lambda$, by boxes $\{B_\varepsilon(\vx_i)\}_i$ of size $\varepsilon$.
From the small scale behavior of $Z(q,\varepsilon)$, one defines the spectrum $\tau_q$ of
scaling exponents:
\begin{equation}
\label{eq:scaling_zq1}
Z(q,\varepsilon)  \operatornamewithlimits{\sim}_{\varepsilon \to 0^+} \varepsilon^{\tau_q} \; .
\end{equation}
It is noteworthy that, from a practical point of view, $\varepsilon \to 0$ means
$\varepsilon \ll L$ where $L$ is some well defined large scale in the problem called
the {\em integral scale}.
Eq. \eqref{eq:def_zq} is thus generally replaced by the following, more stringent, exact scaling relation that amounts to assuming a self-similarity property
of the measure $\Lambda$:
\begin{equation}
\label{eq:scaling_zq2}
Z(q,\varepsilon)  \simeq  Z_q \; \lve^{\tau_q} \; \mbox{for} \; \; 0<\varepsilon \leq L.
\end{equation}
We can remark that, from definition \eqref{eq:def_zq}, we have $\tau_1 = 0$ (by the additivity of the measure) and $\tau_0 = -D_c$ where $D_c$ is the fractal dimension (also called the ``capacity'') of the set $\cS$
that supports the measure $\Lambda$.
In order to characterize the distribution of local singularity exponents $\alpha$,
one introduces the so-called singularity spectrum $f(\alpha)$
defined as the fractal (Hausdorff) dimension of the iso-singulariy sets:
\begin{equation}
\label{deffalpha}
f(\alpha) = Dim_H \{ \vx_0, \alpha(\vx_0) = \alpha \}
\end{equation}
Roughly speaking, this equation means that at scale $\varepsilon$, the number of boxes
where $\Lambda(B_\varepsilon(\vx_i)) \sim \varepsilon^\alpha$ is:
\begin{equation}
N(\varepsilon,\alpha) \sim \varepsilon^{-f(\alpha)} \; .
\end{equation}

According to the multifractal formalism, $f(\alpha)$ and $\tau(q)$, as defined
in resp. Eqs. \eqref{deffalpha} and \eqref{eq:scaling_zq1}, are Legendre transform each other:
\begin{eqnarray*}
	f(\alpha) & = & \min_q(q\alpha-\tau_q) \\
	\tau_q & = & \min_\alpha(q\alpha-f(\alpha))
\end{eqnarray*}

It results that $q$ can be interpreted as a
value of the derivative of  $f(\alpha)$ and conversely
$\alpha$ is a value of the slope of $\tau_q$.
Notice that in the particular case when $\tau_q$ is linear with a slope $\alpha_0$, one recovers the fact that $f(\alpha) = \alpha_0$
for the unique value $\alpha = \tau'_q = \alpha_0$.

Let us emphasize that they are many alternative formulations of the above ``box counting'' method as for instance the correlation integral
approach or the sandbox method (\cite{GP83,federbook}). 
In this paper we will consider, instead of the partition functions
\eqref{eq:def_zq}, the following variant method that involves the 
spatial moments of $\Lambda(B_\varepsilon(\vx))$ over its support:
\begin{equation}
\label{eq:def_mq}
   M(q,\varepsilon) = \langle \Lambda(B_\varepsilon(\vx))^q \rangle = \int G(z,\varepsilon) z^q dz
\end{equation}
where we have denoted by $\langle . \rangle$ the empirical mean over the spatial
support $\cS$ of the measure $\Lambda$ and $G(z,\varepsilon)$ the spatial
distribution\footnote{Notice that in the partition functions \eqref{eq:def_zq}, when $q <0$ the estimate 
	$Z(q,\varepsilon)$ is very unstable because it is  mainly governed by small values of  $\Lambda(B_\varepsilon(\vx_i))$ that are very sensitive to measurement errors or the peculiar choice of the partition ${\cal P}_\varepsilon$. For $q<0$ one generally uses alternative approaches like, e.g., the so-called fixed-mass method (\cite{Mach_1995}) the WTMM method (\cite{wtmm1,wtmm2}) or the more recent Wavelet Leaders method (\cite{leaders1}). However the definition \eqref{eq:def_mq} is more stable when $q<0$ since it involves only boxes centered at points on the support of $d\Lambda$ and thus it is less sensible to the partition choice and to the occurrence of arbitrary close to zero values.} of $\Lambda(B_\varepsilon(\vx))$ at $\vx \in \cS$.
The scaling behavior of $M(q,\varepsilon)$ defines the spectrum of exponents $\zeta_q$:
\begin{equation}
\label{eq:def_zetaq}
M(q,\varepsilon)  \simeq K_q \; \lve^{\zeta_q}
\end{equation}
which can be related to $\tau_q$ as:
\begin{equation}
\label{eq:rel_tauq_zetaq}
\zeta_q = \tau_q+D_c  \; .
\end{equation}
Indeed, since $Z(0,\varepsilon)$ is the total number of boxes at scale $\varepsilon$ needed to cover $\cS$, 
we have
$$
 \langle \Lambda(B_\varepsilon(\vx))^q \rangle \simeq \frac{Z(q,\varepsilon)}{Z(0,\varepsilon)}
$$
which leads \eqref{eq:rel_tauq_zetaq}, thanks to the scaling relationship
\eqref{eq:scaling_zq2} and to the equality $\tau_0 = -D_c$.
Let us remark that if $q=1$, one has simply (because $\tau_1 = 0$ and $\tau_0 = -D_c$)
\begin{equation}
\label{eq:scaling_M1}
  M(1,\varepsilon) \sim \varepsilon^{D_c}
\end{equation}
meaning that the scaling exponent of $M(1,\varepsilon)$ provides a direct estimation
of the fractal dimension of the set $\cS$. Let us mention that some former works have used 
the ``correlation integral'' (\cite{Telesca2005531}) or the ``sandbox method'' 
(\cite{KANEVSKI2017400}) in order to estimate $D_c$. As discussed in \ref{App1},
these estimations can be biased in a multifractal situation and can lead to significantly underestimate the fractal dimension.

Let us finally remark that Eq. \eqref{eq:rel_tauq_zetaq} entails:
\begin{equation}
\label{eq:falpha_zetaq}
  f(\alpha) = D_c+\min_q (q\alpha-\zeta_q)
\end{equation}
that relates the singularity spectrum to the spectrum $\zeta_q$ of the
scaling exponents of the empirical moments of $\Lambda(B_\varepsilon)$.

\subsection{A maximum-likelihood approach to estimate the
multifractal properties of $d\Lambda(\vx)$}
\label{ssec:mll}
In order to obtain the multifractal spectrum of a Cox point process, one thus needs to estimate
the moments of its intensity measure $\Lambda(B_\varepsilon(\vx))$. However, if one observes only one (or few) realization(s) of the inhomogeneous Poisson process associated with a given intensity measure $d\Lambda(\vx)$, the latter is not directly observable. Indeed, one gets, for each $\vx$ and each box size $\varepsilon$, only one or few samples of the
random variables $N(B_\varepsilon(\vx))$ drawn according to the Poisson law
\eqref{eq:PoissonLaw}.
One could use standard statistical inference methods to estimate the full field $d\Lambda(\vx)$
but we prefer to develop a simpler method that consists in considering a parametrization
of the intensity distribution as a sum (or a ``mixture'') of several simple distributions. The observable distribution will be thus a mixture of compound Poisson random variables.
A standard way to estimate the parameters of this mixture in order
to maximize the (log-) likelihood is to use an Expectation-Maximization (EM)
procedure (see e.g. \cite{learningBook}).
To be more specific, let $g(z,\btheta)$ be a family of probability density functions over $\mathbb{R}^{+*}$ of parameters $\btheta$ and let us represent the distribution $G(z,\varepsilon)$ of
$\Lambda(B_\varepsilon(\vx))$ as the following mixture:
\begin{equation}
\label{eq:mixture_g}
G(z,\varepsilon) = \sum_{k=1}^J w_k \; g(z,\btheta_k)
\end{equation}
where the dependence in the scale $\varepsilon$ may rely in all parameters
$J$, $w_k$ and $\btheta_k$.
Let us mention that the weights $\{w_k\}_{k=1,\ldots,J}$ are such that $0 \leq w_k \leq 1$ and $\sum_{k=1}^J w_k = 1$.
A mixture distribution such as $G(z,\epsilon)$ can be interpreted as a weighted sum of conditional distributions corresponding 
to $J$ different ``pure states'' whose probability density functions are $\{g(z,\btheta_k)\}_{k=1,\ldots,J}$. In that respect, $w_k$ can be interpreted as the probability 
of being in state $k$. 
By the definition of $N(d\vx)$ as an inhomogeneous Poisson process
of intensity $d\Lambda(\vx)$, $P(n,\varepsilon)$,
the empirical distribution of observed events $N(B_\varepsilon (\vx))$,
can be written as:
$$
  P(n,\varepsilon) = Prob\left\{N(B_\varepsilon(\vx)=n\right\} = \frac{1}{n!}
  \int_0^\infty e^{-z} z^n G(z,\varepsilon) \; dz
$$
and therefore, by defining:
\begin{equation}
\label{eq:def_h}
   h(n,\btheta) = \frac{1}{n!} \int_0^\infty e^{-z} z^n g(z,\btheta) d z \; .
\end{equation}
we have the finite mixture representation of $P(n,\varepsilon)$:
\begin{equation}
\label{eq:mixture_p}
  P(n,\varepsilon) = \sum_{k=1}^J w_k h(n,\btheta_k)
\end{equation}
which parameters $\{w_k, \btheta_k\}_{k=1,\ldots,J}$ can be estimated using an EM method. The hyper-parameter $J$ can be chosen using a BIC or AIC selection criterion.

Thanks to Eq. \eqref{eq:def_mq}, the partition function $M(q,\varepsilon)$ is then estimated as:
\begin{equation}
\label{eq:moments_mixture}
 M(q,\varepsilon) = \sum_{k=1}^J w_k \int_0^\infty  \! \! z^q g(z,\btheta_k) d z \; .
\end{equation}
In practice, we will consider the distribution $g(z,\btheta)$ to be
either a Gamma distribution or a log-Normal distribution that have both
2 parameters. The corresponding compound Poisson distributions $h(n,\btheta)$ are respectively Ne\-gative Binomial and Poisson-Log-Normal distributions.
Notice that in the latter case, no closed-form formula is available for $h(n,\btheta)$ and a numerical evaluation of the integral \eqref{eq:def_h} has to be performed. On the other hand, the moments $\int z^q g(z,\btheta_k) dz$ always exist for $q \leq 0$ in the log-normal case
while in the Gamma case one must have $q \geq -q_{min}$ where $q_{min}$ corresponds to the minimum value of the shape parameter over all the Gamma functions involved in the mixture. Therefore, in the following, in order to estimate $\zeta_q$ in a range involving positive as well as negative $q$, we have chosen to use the log-normal family while at the same time verifying that for $q \geq 0$ both families lead to the same results.

Let us remark that for integer values of $q$, the moments of $\Lambda(B_\varepsilon(\vx))$ correspond to the factorial moments of $N(B_\varepsilon(\vx))$. For example one has 
$$\langle N(B_\varepsilon(\vx))^2 \rangle = \langle \Lambda(B_\varepsilon(\vx) \rangle +
\langle \Lambda(B_\varepsilon(\vx)^2 \rangle. 
$$
In that respect, unlike $\langle \Lambda(B_\varepsilon(\vx)^q ) \rangle$, the moments $\langle N(B_\varepsilon(\vx))^q \rangle$ are not expected to possess exact scaling properties since, for $q \in \mathbb{N}$, they involve moments of $\Lambda$ at all orders 
less than $q$ (see the discussion in \ref{App2}). 

The approach we propose in this paper is similar to the one formerly developed in \cite{Hwa95,Jie97} in the context of particle physics where the authors proposed to filter out the ``statistical fluctuations'' (i.e. the Poisson random part) by estimating the ``factorial moments of continuous order" that correspond to the standard moments of the intensity measure. For that purpose, they described the distribution of the observed number of events as a mixture of negative binomial distributions.
\begin{center}
	\begin{figure}[h]
		\includegraphics[width=0.95 \textwidth]{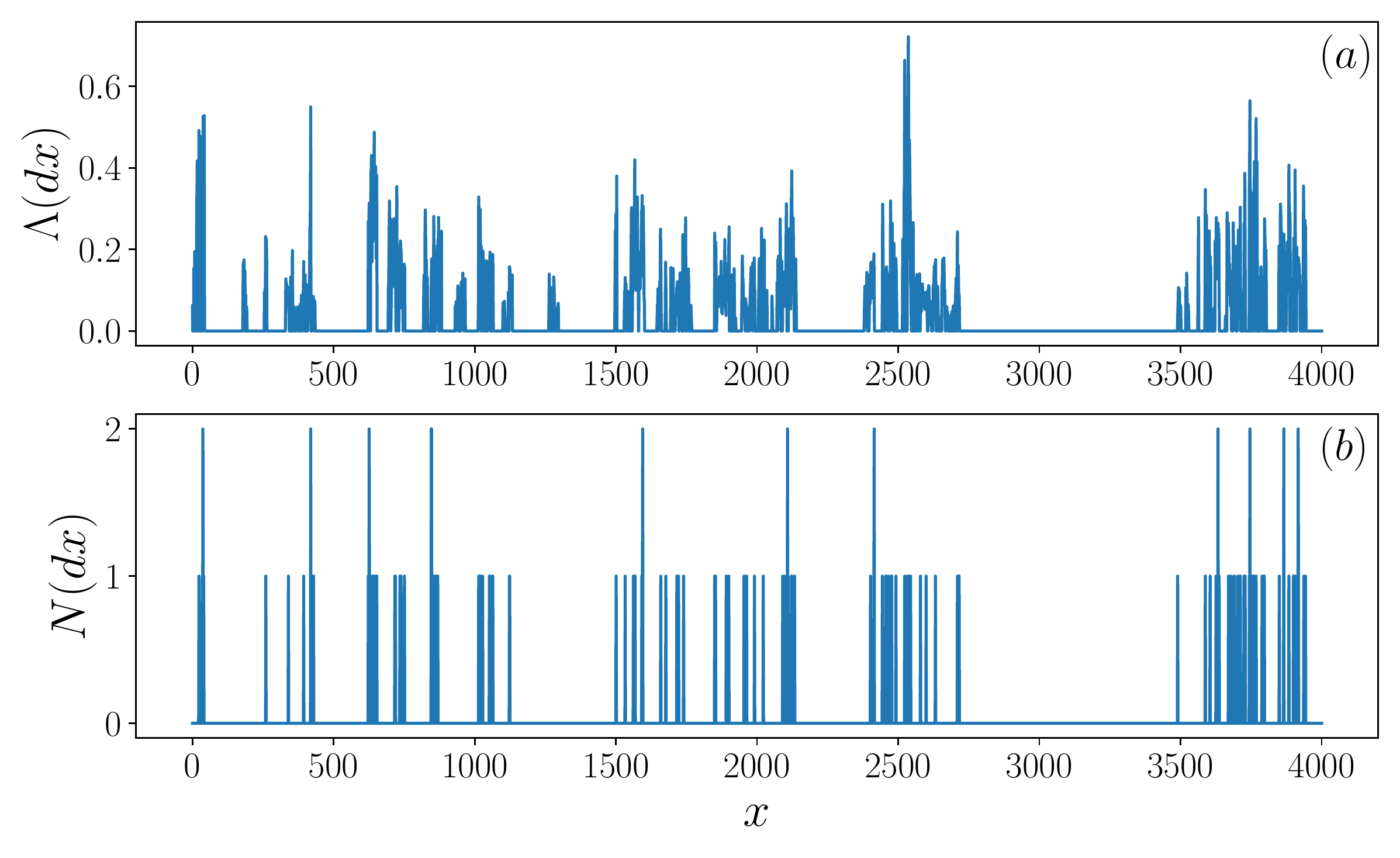}
\vspace*{-0,4cm}  
		\caption{(a) Sample of a log-Normal multifractal measure spread over a Cantor set of dimension $D=0.8$. The integral scale is $L = 512$ and the intermittency
			coefficient $\lambda^2 = 0.05$. (b) Realization of an inhomogeneous Poisson process associated with the multifractal intensity displayed in (a).}
		\label{fig:ex1}
	\end{figure}
\end{center}

\begin{center}
	\begin{figure}[htb]
		\includegraphics[width=1.0 \textwidth]{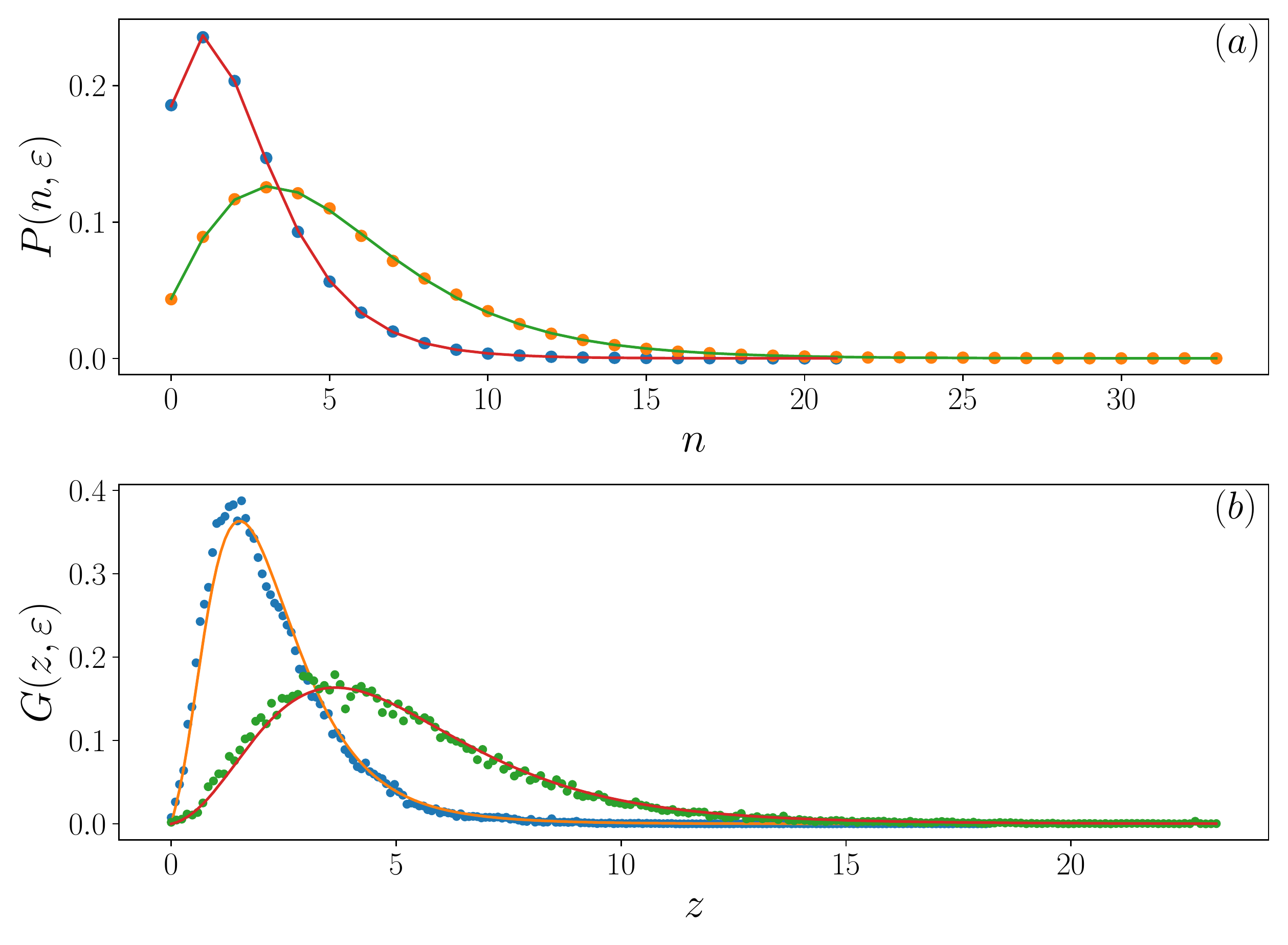}
		\caption{(a) Poisson-Log-Normal mixture (solid line) and empirical ($\bullet$) distribution of $N(B_\varepsilon(x))$ for two scales $\varepsilon = 13$ and $\varepsilon = 40$.  (b) Associated Log-Normal mixture distribution $G(z,\varepsilon)$ as compared to the observed one computed directly from the log-Normal sample of $\Lambda(dx)$. At both scales the mixtures involve $J=3$ log-Normal or Poisson-Log-Normal distributions.}
		\label{fig:fig_ex2}
	\end{figure}
\end{center}

\begin{center}
	\begin{figure}[hbt]
		\includegraphics[width=1 \textwidth]{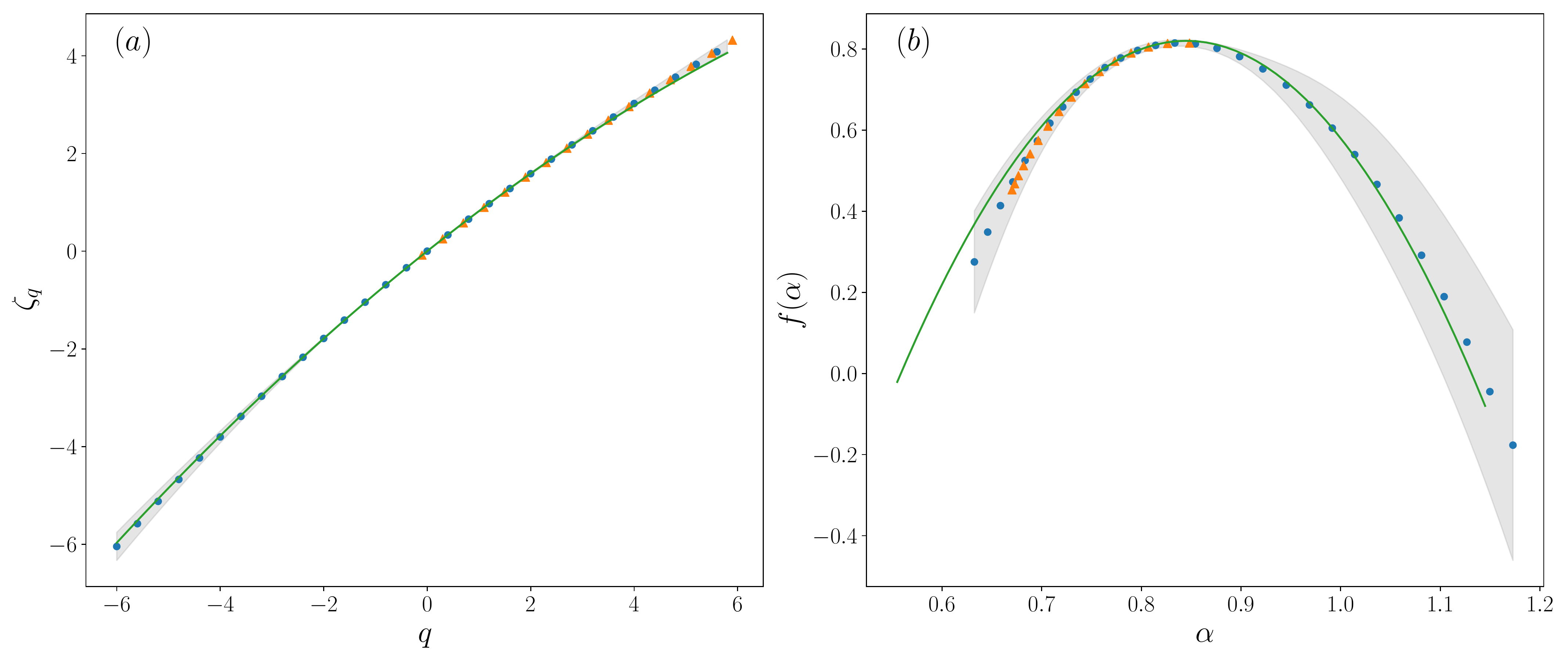}
		\caption{(a) Estimated ($\bullet$) $\zeta_q$ spectrum of moment scaling exponents as compared to the analytical expression \eqref{ln_zeta} (solid line).  (b) Estimated $f(\alpha)$ singularity spectrum ($\bullet$) as obtained by the numerical Legendre transform of the $\zeta_q$ estimation. The solid line represents the expected parabolic spectrum as given by Eq. \eqref{ln_fa}. In both figures, the orange triangles represent the estimations performed with Negative Binomial distributions in the range $q \geq 0$ and the shaded region stands for the observed variations in the estimated spectra when one changes the range of scales to perform the linear fit in log-log representation of the moments.}
		\label{fig:fig_ex3}
	\end{figure}
\end{center}

\subsection{Illustrative examples}
\label{ssec:mll_ex}

\subsubsection{Log-normal multifractal cascade supported by a Cantor set in $\mathbb{R}$.}
Let us consider as a first example, a Cox process in $\mathbb{R}$ ($d=1$) which intensity is a log-infinitely divisible random multiplicative cascade. Such cascade models
represent the paradigm of multifractal processes with well defined
exact scaling properties (\cite{mrw1,mrw2}). More precisely, we study a Cox process with an intensity $d\Lambda(x)$ that is provided by the lacunary random cascade model defined in \cite{MuzyBaile16}. This model consists
in building a log-infinitely divisible random cascade that is supported by a Cantor set of arbitrary dimension $D \leq 1$.
In our example, we consider a measure $\Lambda(dx)$ distributed on a set
of dimension $D = 0.8$ with log-normal multifractal statistics
of intermittency coefficient $\lambda^2 = 0.05$ and integral (i.e. maximum correlation) scale $L = 512$ (see \cite{MuzyBaile16} for
the precise meaning of these parameters).

Such a measure can be shown to be
multifractal with a moment scaling exponents and singularity spectrum
that are quadratic functions:
\begin{eqnarray}
\label{ln_zeta}
	 \zeta_q  & = & \alpha_0 q -\frac{\lambda^2}{2}q^2  \\
\label{ln_fa}
	  f(\alpha) & = & D-\frac{(\alpha-\alpha_0)^2}{2 \lambda^2} \; \mbox{with} \; \;
	   \alpha_0 =  D+\frac{\lambda^2}{2}
\end{eqnarray}

A sample of $\Lambda(dx)$
(where we choose numerically $dx = 1$)  over 8 integral scales is represented in Fig. \ref{fig:ex1}(a). A realization of the associated counting process $N(dx)$ (each interval $dx$ contains a random number drawn with a Poisson law of intensity $\Lambda(dx)$) is displayed in Fig. \ref{fig:ex1}(b). The latter process can be viewed as a ``noisy'' version of $\Lambda(dx)$ where Poisson statistical fluctuations are superimposed to the spatial log-normal cascade noise. All data have been generated by a numerical simulation of the model as described in \cite{MuzyBaile16}.
If one wants to characterize the genuine ``risk'', i.e., the statistical properties of $\Lambda(dx)$, one has to get rid of these Poisson fluctuations. This is the purpose of the previously described maximum likelihood method relying on representations \eqref{eq:mixture_g}, \eqref{eq:mixture_p}
that aims at recovering $G(z,\varepsilon)$, the distribution of $\Lambda(B_\varepsilon(x))$
from $P(n,\varepsilon)$, the distribution of $N(B_\varepsilon(x))$.
We choose to represent $P(n,\varepsilon)$ as a mixture of $J$ Poisson-Log-Normal 
distributions which amounts to representing $G(z,\varepsilon)$ as a
mixture of $J$ Log-Normal distributions. 
In the range $q>0$, we checked that we recover the same results using negative binomial distributions (represented by orange triangles in Fig. \ref{fig:fig_ex3}).
We performed our numerical estimation using a sample $N(dx)$
over a total length of $256$ integral scales.
Our analysis was performed over a range of scales $\varepsilon$ such that $1 \leq
\varepsilon \leq 100$. At each scale, the parameter $J$ can be determined using a BIC or AIC selection criterion. 
We find that, for all scales, $J=3$
achieves (or almost achieves) the maximization of the penalized likelihood.


The performance of the parametric maximum likelihood method to fit the empirical
distributions of $N(B_\varepsilon(x))$ at each scale is illustrated in Fig.
\ref{fig:fig_ex2} for scales $\varepsilon = 13$ and $\varepsilon = 40$.
One can see that in both cases, the observed empirical distributions are fitted
fairly well by a Poisson-Log-Normal mixture with $J=3$. It results that $G(z,\varepsilon)$, the original distributions of $\Lambda(B_\varepsilon(x))$, are also well fitted by the associated mixtures of Log-Normal distributions (Fig.
\ref{fig:fig_ex2}(b)). From these mixtures, $M(q,\varepsilon)$ is then computed
at each scale $\varepsilon$ using Eq. \eqref{eq:def_mq} and from a linear fit of $\ln(M(q,\varepsilon))$ as a function of $\ln \varepsilon$, one gets an estimate of the spectrum $\zeta_q$. The estimation obtained from the sample of length
$256$ $L$ in our example is reported in Fig. \ref{fig:fig_ex3}(a). The shaded grey
region indicates an order of magnitude of the estimation error when one changes
the range of scales used to perform the fit from smallest scales to largest ones.
We can see that the numerical estimation procedure detailed previously allows one to
recover precisely the expected parabolic $\zeta_q$ function.
It results, using a numerical Legendre transform that one can estimate also
quite well the shape of the singularity spectrum for this model (Fig. \ref{fig:fig_ex3}(b)).

\subsubsection{Monofractal examples}
In a second example, we consider various mono-fractal situations when $d=2$. 
We construct random Cantor sets
of dimension $D_c < 2$ in $\mathbb{R}^2$ that are the two-dimensional versions of the random Cantor sets introduced in \cite{MuzyBaile16} whose law is invariant by space translations and possesses exact self-similarity properties.
It results that if we consider a measure $d \Lambda$ uniformly spread on such sets, we have:

\begin{eqnarray}
\label{mono_zeta}
\zeta_q  & = & D_c \; q  \\
\label{mono_fa}
f(\alpha) & = & 
\begin{cases}
D_c  \; \; & \mbox{if} \; \; 
\alpha =  D_c  \\
-\infty  \; & \mbox{otherwise.} 
\end{cases}
\end{eqnarray}
\begin{center}
	\begin{figure}[h]
		\hspace*{0.5cm}
		\includegraphics[width=1.0 \textwidth]{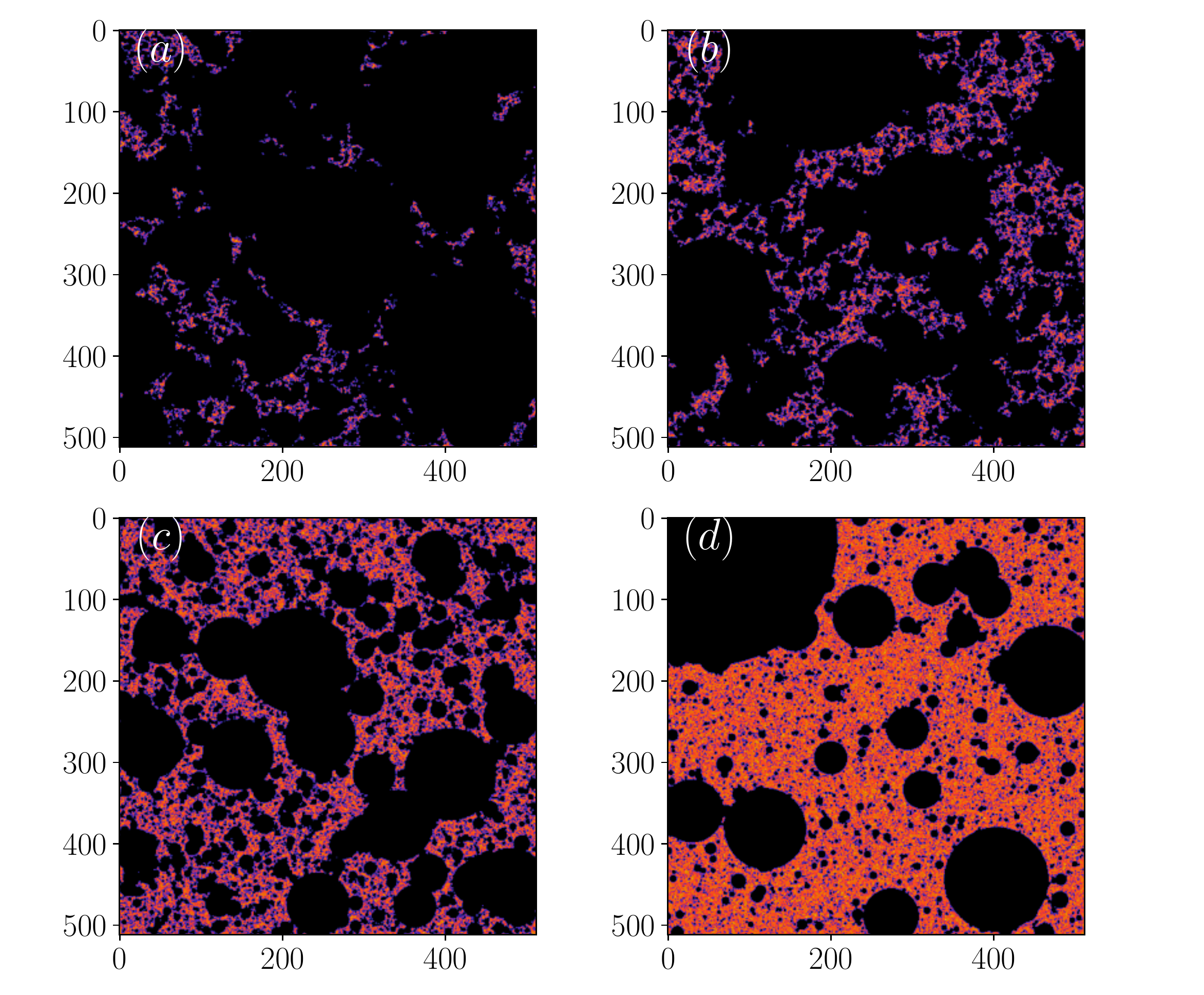}	
		\caption{Four examples of uniform intensity spread on a random (Cantor) set. Each realization is sampled on a $512 \times 512$
			grid. They correspond to respectively to dimensions (a) $D_c = 1.3$, (b) $D_c=1.5$, (c) $D_c = 1.7$ and (d) $D_c = 1.9$.}
		\label{fig:ex2_1}
	\end{figure}
\end{center}
\begin{center}
	\begin{figure}[h]
		\includegraphics[width=1 \textwidth]{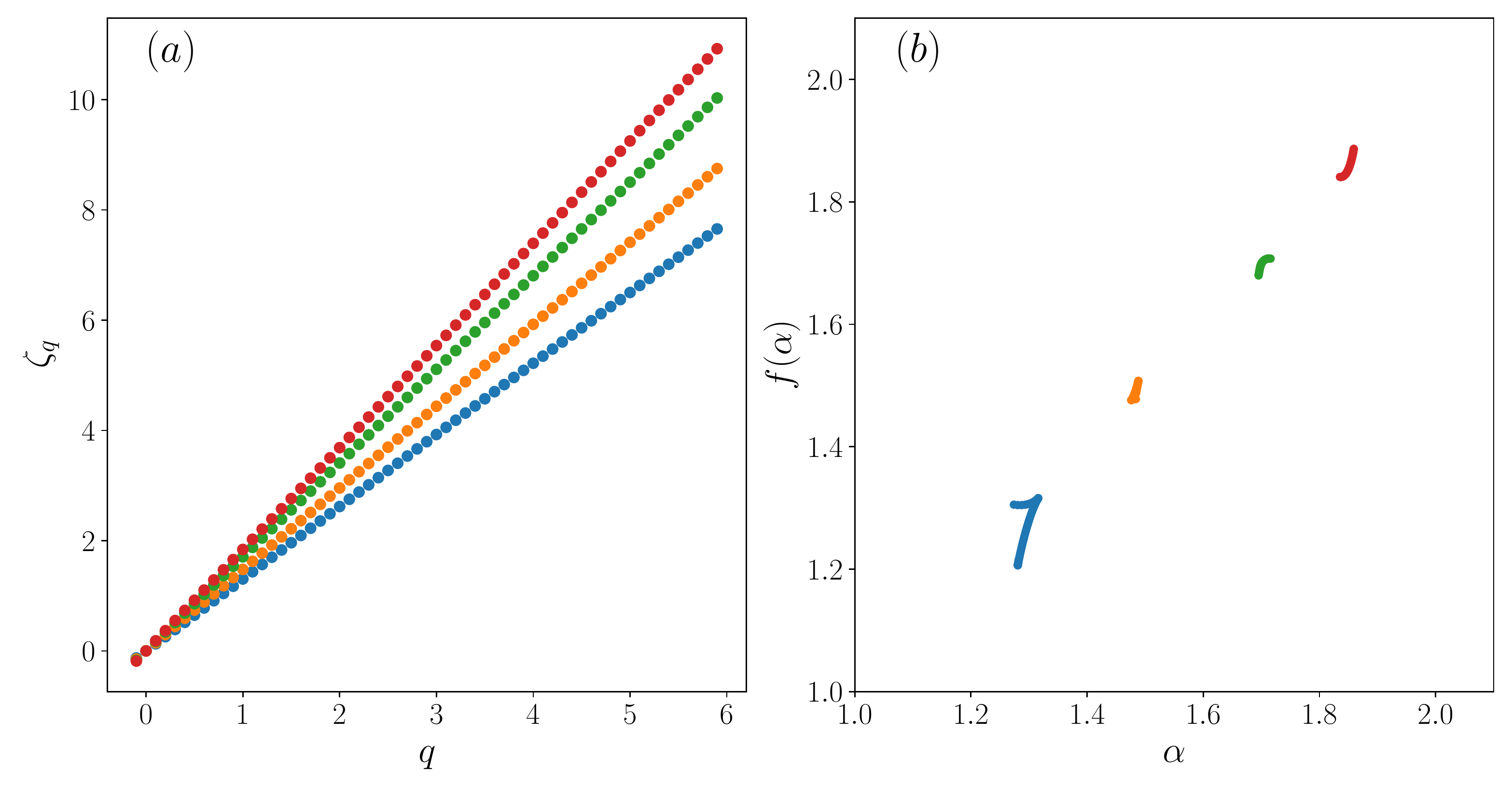}
		\caption{(a) Estimated $\zeta_q$ functions (for $q \geq 0$) of the four examples diplayed in Fig. \ref{fig:ex2_1}. On clearly sees that each curve is almost linear 
			with a slope provided by the dimension of the support in perfect agreement with Eq.	\eqref{mono_zeta}
			(b) Estimated singularity spectra obtained from numerical Legendre transform. In each case, up to small deviations observed at large $q$, one recovers the analytical expression \eqref{mono_fa}.}
		\label{fig:ex2_2}
	\end{figure}
\end{center}

\vspace*{-1cm}
In that respect, such mono-fractal random measures can be considered as the analog
of fractional Brownian motion for random functions.

In Fig \ref{fig:ex2_1} are displayed four examples (of dimensions $D_c = 1.3,1.5,1.7$ and $1.9$) of these random Cantors built on a $512 \times 512$ grid. More precisely we have reported, at each pixel, a color corresponding to the observed number of events of a Cox process associated with a uniform random measure spread over each of these sets. As illustrated in Fig \ref{fig:ex2_2},
when we estimate $\zeta_q$ using the method described in Sec. \ref{ssec:mll} with Poisson-Log-Normal mixtures, we recover quite well the expected theoretical expressions provided in Eqs. (\ref{mono_zeta}, \ref{mono_fa}) namely a
straight line of slope $D_c$ for $\zeta_q$ and $f(\alpha)$ spectrum that reduces (within statistical errors) to one point $(D_c,D_c)$ in each case.

\section{The multifractal approach applied to the Prom\'{e}th\'{e}e database }
\label{sec:multifracapp}

\subsection{The Prom\'{e}th\'{e}e database and the annual ignition intensity measure}
\label{sec:m1}
The Prom\'{e}th\'{e}e database (www.promethee.com) was created in 1973 in order to gather several informations relative to the wildfire in French Mediterranean regions. It reports many different features of the wildfire occurred in 15 French departments (year, region administrative number, fire number, geographical coordinates, date, burnt surface, characteristics of the first fire fight action, nature of the damages, vegetal species, information relative to the fire cause, ...). The information flux sources come from various national services acting in each region (fire fighters, forest managers, police, civil safety, army and air force). Some efforts to homogenize these multiple data were performed in the 80's, especially through a new system of coordinates specially designed for fire management: the DFCI coordinates (DFCI is the French acronym for  "D\'efense de la For\^et Contre les Incendies", i.e. it refers to all processes concerning forest defense facing to wildfires). This coordinate system consists in a set of nested grid layers with increasing resolution going for $100 \times 100$ $km^2$ to $2 \times 2$ $km^2$ at the finest resolution.
The fire ignition locations are therefore available with the rather poor spatial resolution of $4$ $km^2$.
Let us mention that Prom\'eth\'ee data have already been used in various academic works devoted to the empirical study of wildland fire hazards (see for example \cite{Mang06,GANT13,Ager2014,Lahaye2014,Opitz2017,Opitz2020}).

In order to handle consistent data and to avoid biased results, we choose to process the database restricted from the 1$^{st}$ of January, 1992 to the 31$^{st}$ of December, 2018.
We consider separately three main regions namely, ``Corsica", ``Provence-Alpes-C\^ote d'Azur'' (PACA) and ``Languedoc-Roussillon'' (LR). The main statistics of the database are summarized in table \ref{Promregion} where we see that the number of reported ignitions is larger in Corsica than in the two
other regions whilst it is almost four times less wide.

\begin{table}[h]
	\begin{center}
    \begin{tabular}{ | c || c | c | c | c | c | c |}
     \hline			
       Region & Surface ($km^2$) & $N_{tot}$ & $N_{S>1000}$ & $N_{S>10000}$ & $N_{S>50000}$    \\
       \hline			
       Corsica &  8680  & 21073  &  8872   & 4666 &   1449 \\
       \hline
       PACA & 31400  & 18213 &  10734  &  3472 & 1151 \\
       \hline
        LR &  27376 & 12552 & 9155 & 5143 &  1396 \\

       \hline
 \end{tabular}
\end{center}
 \label{Promregion}
       \caption{Main statistical features of the Prom\'eth\'ee database: we reported the overall surface of each regions,
       the number of forest fire events observed from 1992 to 2018,
       and the number of the events where the burnt area is greater than respectively
       1000,10000 and 50000 $m^2$.}
\end{table}

\begin{figure}[t]
	\hspace*{-1.2cm}
	\vspace*{-0.5cm}
	\includegraphics[width=18cm]{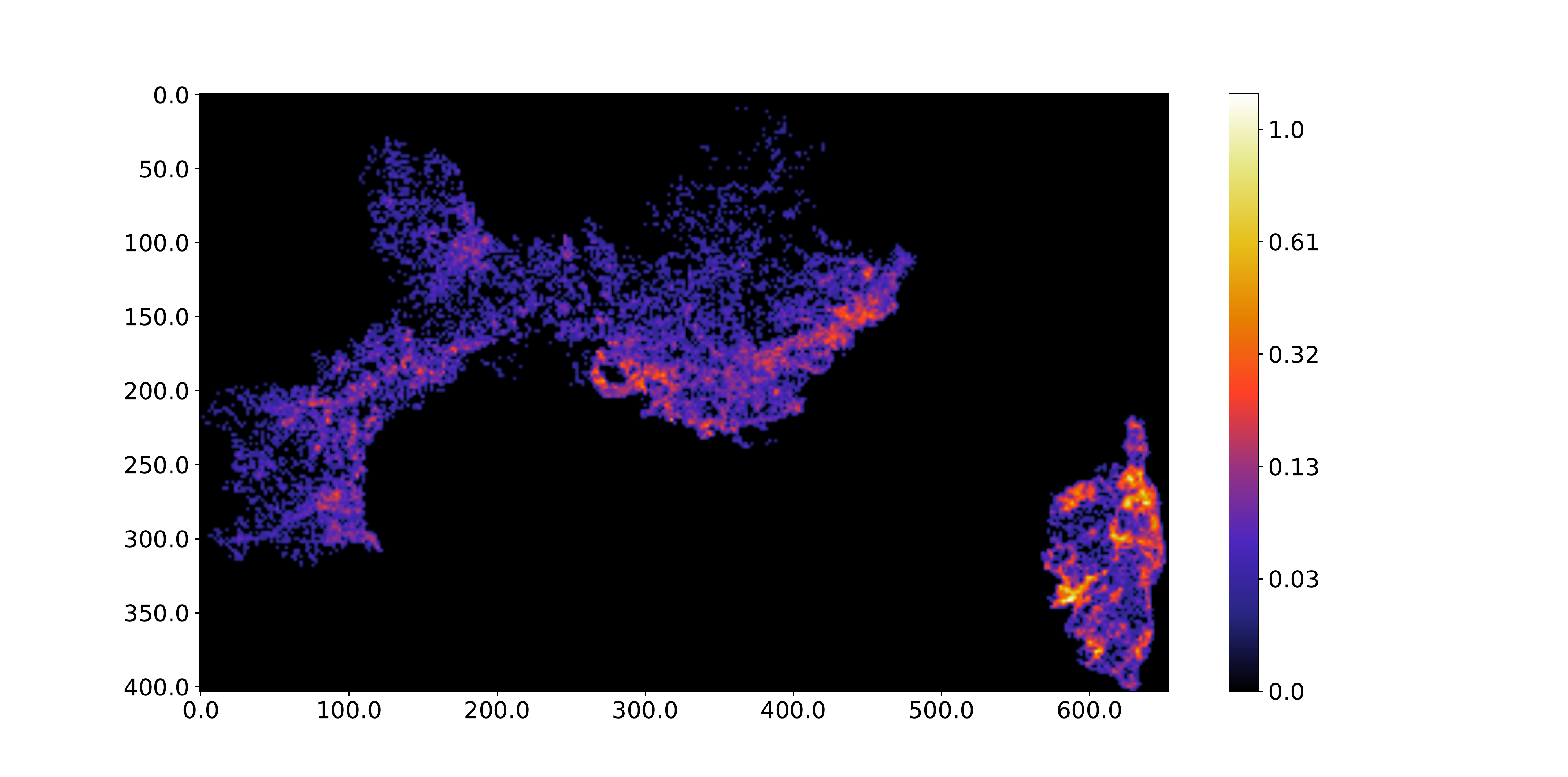}
	\caption{Map $\tL_k$ of annual fire occurrence rate for the Southern French Mediteranean regions as estimated from Prom\'{e}th\'{e}e database over the period 1992-2018. The axis origin has been chosen arbitrarily and the reported spatial scales are in $km$ on both axis.}
	\label{cor_vs_pyor}
	\label{fig:density}
\end{figure}

The fire ignition process over a given region can be considered as a
time-space Point process and therefore the intensity (or expectation) measure
$d\Lambda(\vec x,t)$, i.e., the mean number of events between $t$ and $t+dt$ in a square of size $d^2 x$ located at $\vx$,  depends on both $\vx$ and $t$.
Notably, as already discussed in former studies (see e.g. \cite{Zhang2014,BAJOCCO2017433}), $\Lambda(d^2x,dt)$ displays strong annual seasonality, the ignition rate being much larger in the dry summer
season than in the winter season.  Let us mention that \cite{Opitz2017,Opitz2020} tried to precisely
capture the space-time variations of $d\Lambda(\vec x,t)$ and its relationship with exogenous "covariates" like climatic variables through the calibration of a spatio-temporal log-Gaussian Cox model.
Since our goal is to study the statistical properties of the spatial fluctuations of the fire occurrence likelihood and mainly its scaling properties, in order to avoid these
seasonal effects and to consider a purely spatial Point process, we focus in the paper on the annual ignition rate.  Hereafter, $d\Lambda(x,y)$ will stand for the intensity
associated with the annual number of events at some given location $\vx=(x,y)$
over an infinitesimal square of size $dxdy$.
Let us denote by ${\cal S}$ the support of $d\Lambda$, namely the set where $d\Lambda(x,y)$ is non vanishing. It can be empirically defined
as the set of centers $\vx_k$ of the 4 $km^2$ DFCI cells (the smallest available resolution) where there has been at least one event during the whole sample period 1992-2018. Let $S$ be the cardinal
of $\cS$ and let us denote by $\{\Lambda_k\}_{k=1,\ldots,S}$ the intensity
associated with DFCI cell in ${\cal S}$, i.e., $\Lambda_k = \Lambda(B_2(\vx_k))$.
Along the same line, we will denote by $N_k$ the (random) number of events in cell $k$
during a year: $N_k = N(B_2(\vx_k))$. 
In practice, a surrogate for $\Lambda_k$ can be obtained as the
mean number of events per year observed during the whole 27 years period:
\begin{equation}
\label{eq:proxy}
 \tL_k = \frac{1}{27 }\sum_{y=1992}^{2018} N_k(y) \equiv E(N_k)
\end{equation}
where we have denoted by $N_k(y)$ the realization of $N_k$ at year $y$ and defined the expectation $E(.)$ as the average across all 27 years of the sample.
In Figure \ref{fig:density}, we reported such an annual fire occurrence mean density
observed in each DFCI square in the French southern regions.
Density levels are indicated in number of ignitions per $km^2$ and per year. We see that the distribution appears very inhomogeneous in space,
with obvious clustering properties, high ignition levels being observed close to main road axes and densely populated areas.

In order to describe the spatial fluctuations of the annual
number of fire occurrences $dN(x,y)$, we will use the approach described in Sec. \ref{sec:mpp}, where we suppose that $dN(x,y)$ is a Cox process, i.e., conditionally to random multifractal spatial intensity $d\Lambda(x,y)$, $dN(x,y)$ is an inhomogeneous Poisson process.
This assumption notably implies that, for any given intensity
function, the observed number of ignitions during a given year over distinct areas are uncorrelated. This means that the observed spatial clustering of ignition locations 
is exclusively due to the spatial fluctuations of the intensity field.
In \ref{clusterProp}, using the same approach as in \cite{Hering2009} based on Ripley inhomogeneous $L$ function, we show that this assumption is well verified.

\begin{center}
	\begin{figure}[t]
		\hspace*{-0.5cm}
		\includegraphics[width=14cm]{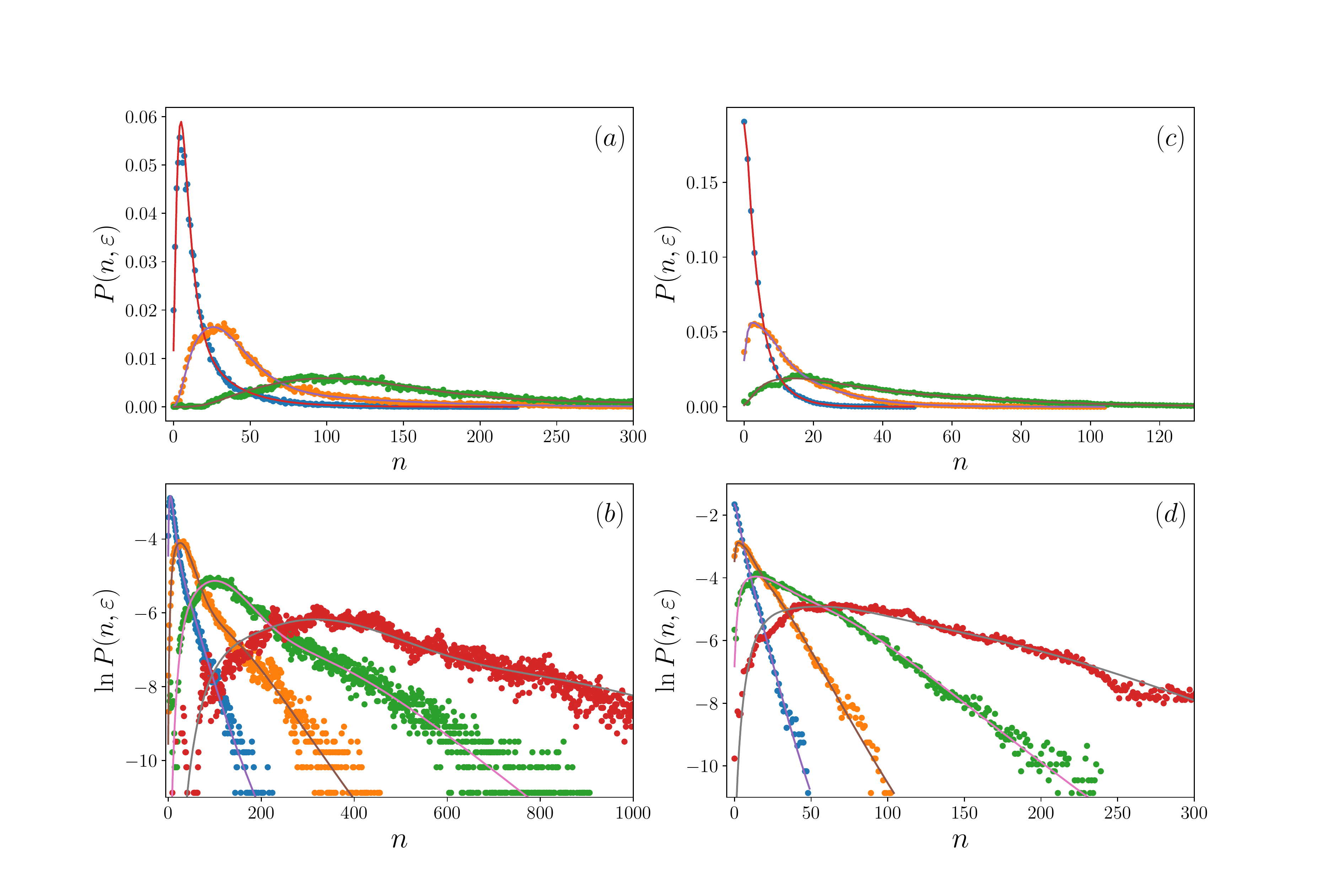}
		\vspace*{-0.5cm}
		\caption{Spatial probability distributions of $N(B_\varepsilon(\vx))$ for $\varepsilon = 6$ (blue points),$12$ (orange points), $22$ km (green points) and $38$ km (red points) in (b,d) in Corsica and LR regions. Panels (a) and (c) are in linear scales while (b) and (d) are in logarithmic scale in order to emphasize the tails of the distributions. The left plots (a,b) correspond to estimates obtained when accounting for all forest fire events in Corsica and the right plots (c,d) correspond to fire events of same type in LR region.  Symbols ($\bullet$) represent the empirical data and the solid lines are the best fit obtained with a mixture of $J=3$ Poisson-Log-Normal distributions.}
		\label{fig:mf_c1}
	\end{figure}
\end{center}

\begin{center}
	\begin{figure}[t]
		\hspace*{-1cm}
		\includegraphics[width=15cm]{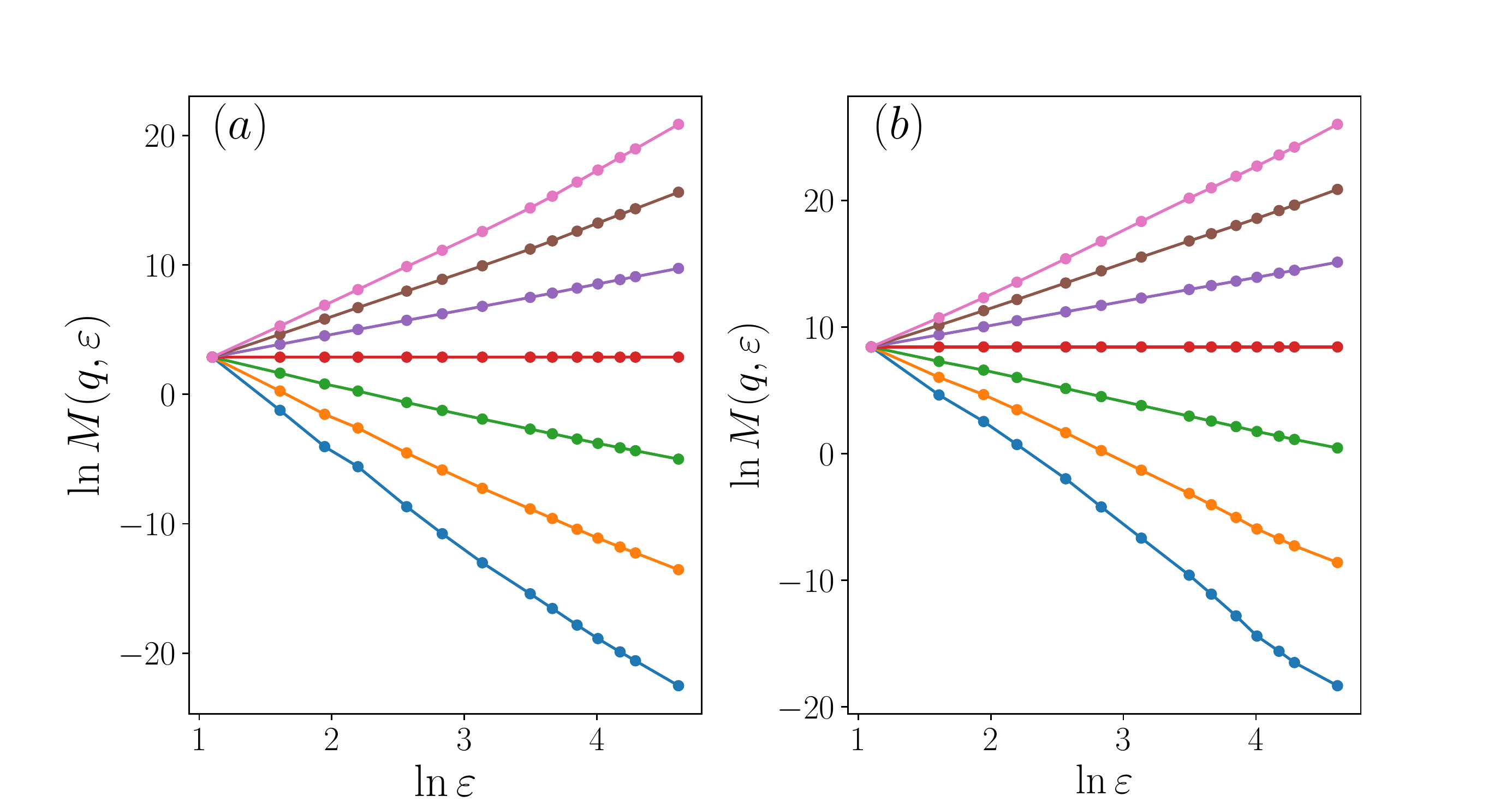}
		\caption{Scaling properties of intensity moments $M(q,\varepsilon$). $\ln M(q,\varepsilon)$, as defined in Eq. \eqref{eq:def_mq}, is plotted as a function
		of $\ln \varepsilon$ for $q=-3,-2,\ldots,3$ (from bottom to top) for all forest fire events from 1992 to 2018 in (a) in Corsica and (b) in LR region.}
		\label{fig:logzq}
	\end{figure}
\end{center}

\begin{center}
	\begin{figure}[t]
		\hspace*{-0.5cm}
		\includegraphics[width=14cm]{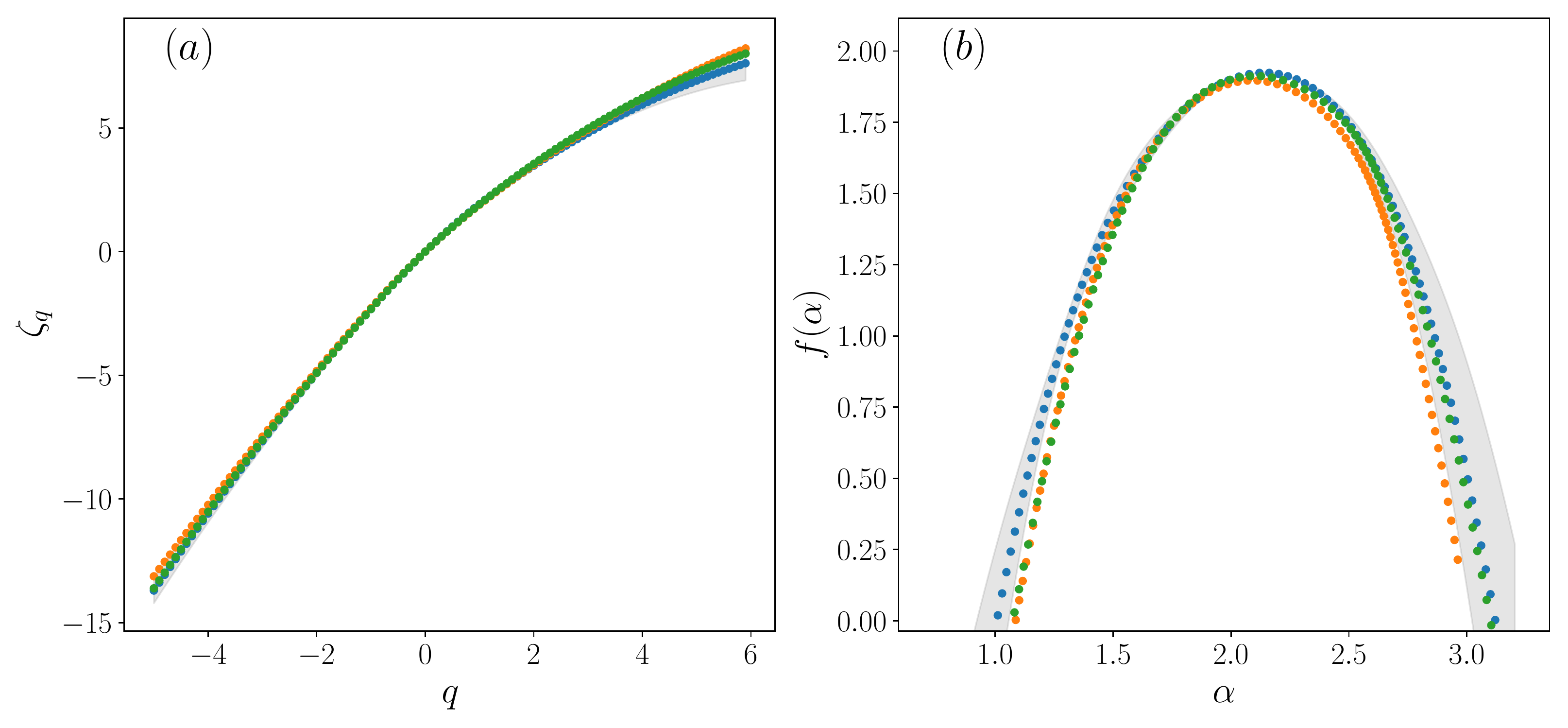}
		\caption{Multifractal spectra $\zeta_q$ and $f(\alpha)$ for the three regions
		as obtained from the empirical scaling properties of $M(q,\varepsilon)$. The grey
	shaded regions correspond to the observed fluctuations in spectra for Corsica data when one changes the fitting interval towards large and small scales. Blue, orange and green symbols correspond respectively to Corsica, LR and PACA regions.}
		\label{fig:spectraAll}
	\end{figure}
\end{center}

\begin{center}
	\begin{figure}[t]
		\hspace*{0.5cm}
		\includegraphics[width=12cm]{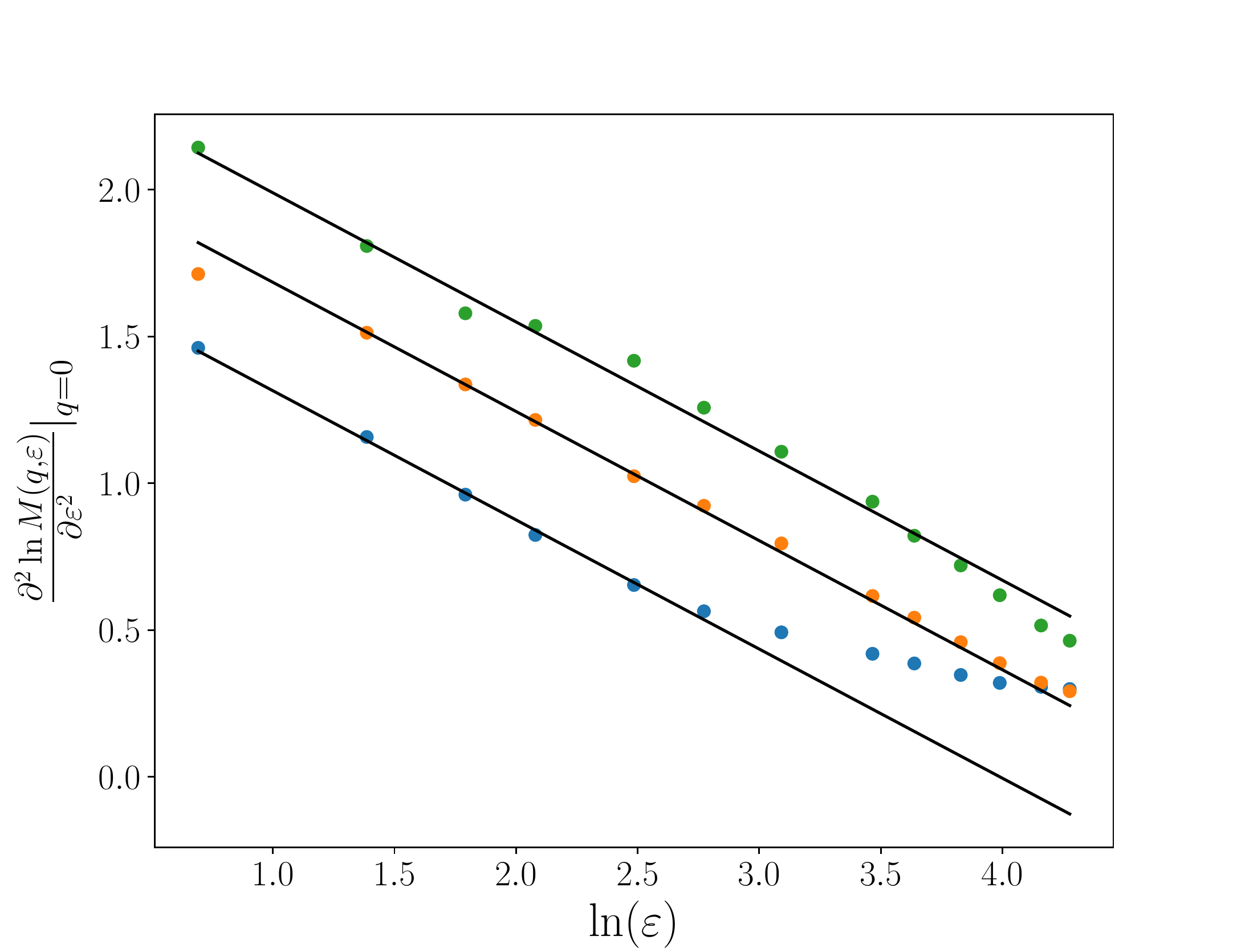}
		\caption{Estimating the intermittency coefficient of the intensity of distribution. $\frac{\partial^2 \ln M(q,\varepsilon)}{\partial q^2} {\big \rvert_{q=0}}   $ is plotted as a function of the scale logarithm $\ln \varepsilon$ for Corsica (blue), LR (orange) and PACA (green) regions. The linear decreasing behavior provides, according to Eq. \eqref{eq:l2_est} an estimation $\lambda^2 \simeq 0.4$ in all cases.}
		\label{fig:l2_est}
	\end{figure}
\end{center}

\subsection{The multifractal nature of wildfire ignition risk distribution}
In this section we report the empirical results we obtained using the data of the 3 regions when estimating the law of $\Lambda(B_\varepsilon(\vx))$
and then the multifractal scaling laws $\zeta_q$, $f(\alpha)$ following the
method of Sec. \ref{ssec:mll}.
We chose to compute the probability distribution at points $\{\vx_k \}_{k=1,\ldots,S}$
that are the centers of the DFCI
cells in the set ${\cal S}$ where there has been at least one event.
Let us mention that, along the same line than in \ref{clusterProp}, we have taken into account the edge effects by applying a Ripley edge correction factor $w_{ij}$ when estimating, for each $\vx_k$, $N(B_\varepsilon(\vx_k))$ from observations
at smallest scale ($\varepsilon = 2$) $N_j$:
$$
N(B_\varepsilon(\vx_k)) = \sum_{\vx_j \in B_\varepsilon(\vx_k)} w_{kj}^{-1} N_j
$$

We estimate the empirical distributions of $N(B_\varepsilon(\vx_k))$ for
$\varepsilon$ in the range $[2,72]$ $km$ of each region for all reported forest
fire events. For each scale, we compute the best
representation of the distribution in terms of a mixture Negative Binomial
or Poisson-Log-normal distributions. Since both approaches lead to similar results when $q \geq 0$,
we only report the results relative to Log-Normal distributions that are harder to
handle from a numerical point of view but provide stable results in the range $q < 0$. 
As discussed previously, the hyper-parameter $J$ can be chosen, at each scale, by the mean of a BIC selection criterion. We found that in all cases $J=3$ provides a near optimum choice for all scales.

In Fig. \ref{fig:mf_c1}, we have displayed the empirical distributions and their
fit using a Poisson-Log-Normal mixture, of annual
number of fire occurrences in boxes of different sizes in Corsica and Languedoc Roussillon regions (plots for the PACA region are similar).
We can see that the shape of these distributions strongly depends on the considered
scale and appears to be over-dispersed as respect to a simple Poisson law.
In both linear (Figs.\ref{fig:mf_c1}(a,c)) and logarithmic plots (Figs.\ref{fig:mf_c1}(b,d)), we can also see that the mixture model provides
a very good fit of the empirical data, around their maximum values as well as in their
tail behavior. 
From Eq. \eqref{eq:def_mq}, one can thus compute the $q$ order moments of the intensity measure at all scales $\varepsilon$ and its scaling exponent $\zeta_q$. In Fig. \ref{fig:logzq}, the estimated $M(q,\varepsilon)$
are displayed in log-log representation for $q = -3,2,\ldots,3$ for
the spatial intensity of all wildland fire events in Corsica and LR regions. We see that in both cases, the scaling assumption
\eqref{eq:def_zq} is sound since $M(q,\varepsilon)$ is rather well modeled by power-law
over a range extending up to the largest scales.

The values of $\zeta_q$ and $f(\alpha)$ spectra we estimated for each of the 3 regions of the database are reported in Fig. \ref{fig:spectraAll}. One can see that all estimated spectra have a well pronounced strictly concave shape specific of a multifractal measure.
Moreover, these empirical functions are very close to each other indicating a somehow universal character of the multifractal properties of $d\Lambda(\vx)$ in the French Mediterranean regions. We notably find that for all 3 distributions, the forest fire ignitions occur on a set of dimension $D_c \simeq 1.9$ with a local exponent $\alpha_0 \simeq 2$. This fractal dimension of the support of the ignition intensity, close to $D=2$, is greater than the values $D_c \simeq 1.5$
reported  in the literature for north Italy (\cite{TUIA20083271,TELESCA20071326}) 
but closer to $D_c \simeq 1.7$ reported in Portugal (\cite{KANEVSKI2017400}). These discrepancies 
may indicate that the geometries of wildfire event supports in different regions are genuinely different but may also be explained by our argument developed in \ref{App1}: estimators relying 
on second order properties like sandbox method or the correlation integral method used in the previously cited works, provide a biased (underestimated) measure of the fractal dimension for a multifractal point pattern.
Moreover, as notably discussed by \cite{Tuia2008}, there can be
several additional statistical biasing factors like the overall number of considered events or the boundary effects.  

It is noteworthy that the spectrum of local singularity exponents extends over an interval $\alpha \in [1.0,3.1]$ for all three regions. It is tempting to interpret the extreme value $\alpha = 1$ 
as being associated with events uniformly occurring along linear geometries 
while the most probable values $\alpha = 2$ could be associated with simple two dimensional domains. Such interpretation remains to be confirmed by a study focusing on local properties that would probably request better resolved spatial data. Let us notice that the range of observed local 
scaling exponent values $\alpha \in [1,3]$ is remarkably agreeing with the range of 
``local dimensions'' observed by \cite{KANEVSKI2017400} using a local sandbox method. 

The width of the range of observed singularity exponent values, $\Delta \alpha$ is often used as a measure of the multifractality strengh. For the three regions, we observe consistently\footnote{Recall that estimation errors we consider correspond to the variations of measured scaling exponents when one changes the scaling range. They are not errors due to statistical noise that are quite hard to determine.} $\Delta \alpha = 2.1 \pm 0.1$.  
Let us notice however, that if $\Delta \alpha = \alpha_{\max}-\alpha_{\min}$ is a measure of the multifractality strength that is well suited to deterministic situations (i.e. involving deterministic multifractal functions), it is hardly adapted to random situations. Indeed,  
the values $\alpha_{\max}$ and $\alpha_{\min}$ involve the computation of scaling exponents of respectively extreme negative and positive moments, which are by far less reliable than those of lowest $q$ values. This is why one generally prefers to estimate the so-called {\em intermittency coefficient} $\lambda^2$ originally introduced in the context of fully developed turbulence (\cite{Fri95}). It quantifies the non-linear character of $\zeta_q$ 
as its curvature at $q=0$ (see. e.g., \cite{Fri95,MuzyBaile16}):
\begin{equation}
\lambda^2 = -\zeta''_0 = -\frac{\partial^2 \zeta_q  }{\partial q^2} {\biggr \rvert_{q=0}}  \; .
\end{equation}
For a log-normal cascade, as in the example we considered in Sec. \ref{sec:multifracapp}, 
when the dimension $D_c$ of the support is fixed, $\lambda^2$  fully
characterizes the shape of the multifractal spectra  {(Eqs \eqref{ln_zeta}, \eqref{ln_fa}). 
One has notably, in that case $\Delta \alpha = 2 \sqrt{D_c \lambda^2}$.
In order to estimate $\lambda^2$, one can show, from Eq. \eqref{eq:def_zetaq} that
\begin{equation}
\label{eq:l2_est}
 \frac{\partial^2  \ln M(q,\varepsilon)}{\partial q^2} {\biggr \rvert_{q=0}} \! \! = -\lambda^2 \ln \lve + V_0
	\end{equation}
with $V_0 = \frac{\partial^2 \ln K_q}{\partial q^2} {\big \rvert_{q=0}} $.
This equation can be interpreted quite easily within a multiplicative
cascade picture. Indeed, $\frac{\partial^2 \ln M(q,\varepsilon)}{\partial q^2} {\big \rvert_{q=0}}  $ is nothing but the variance of $\ln \Lambda(B_\varepsilon(\vx))$ as respect to its
spatial fluctuations. When one goes from fine to large scales, this variance decreases
as linear function of $\ln(\varepsilon)$, meaning that each time one divides the resolution by, e.g., a factor 2, one adds a random term $\omega$ of constant variance to $\ln \Lambda (B_\varepsilon(\vx))$, i.e. the value of $\Lambda(B_\varepsilon(\vx))$ is multiplied by a random factor $W=e^\omega$ and thus it is a random multiplicative cascade. The slope of this linear function, i.e., the quantity $\lambda^2 \ln(2)$,
thus corresponds to the variance of $\omega = \ln W$.
Notice that within this picture, $V_0$ is interpreted as the large scale variance of $\ln  \Lambda(B_\varepsilon(\vx))$ that, when $\varepsilon \geq L$, no
longer depends on $\varepsilon$.

Eq. \eqref{eq:l2_est} provides a simple way to directly estimating $\lambda^2$ from empirical
data. In Fig. \ref{fig:l2_est} are reported the second order derivatives of $\ln M(q,\varepsilon)$
around $q=0$ (as approximated by a finite difference scheme with $\Delta q = 0.2$) for Corsica,
LR and PACA regions. The three curves appear strikingly to be parallel\footnote{This is true only at small scales in the case of Corsica. One can suppose that the large scale behavior can be biased by boundary effects since Corsica region is the narrowest of the 3 considered regions. Moreover, as confirmed on numerical experiments on exact multifractal models, finite sample fluctuations can be quite important on such curves.} with a slope $\lambda^2 \simeq 0.4$ .  This means that the 3 spatial distributions of fire ignitions can be described
by the same cascade model. The precise value of the integral scale $L$ is hard to estimate without
the knowledge of $V_0$ but one can evaluate the magnitude order of their ratio (provided $V_0$ remains
constant across regions).

\begin{center}
	\begin{figure}[t]
		\hspace*{0.5cm}
		\includegraphics[width=12cm]{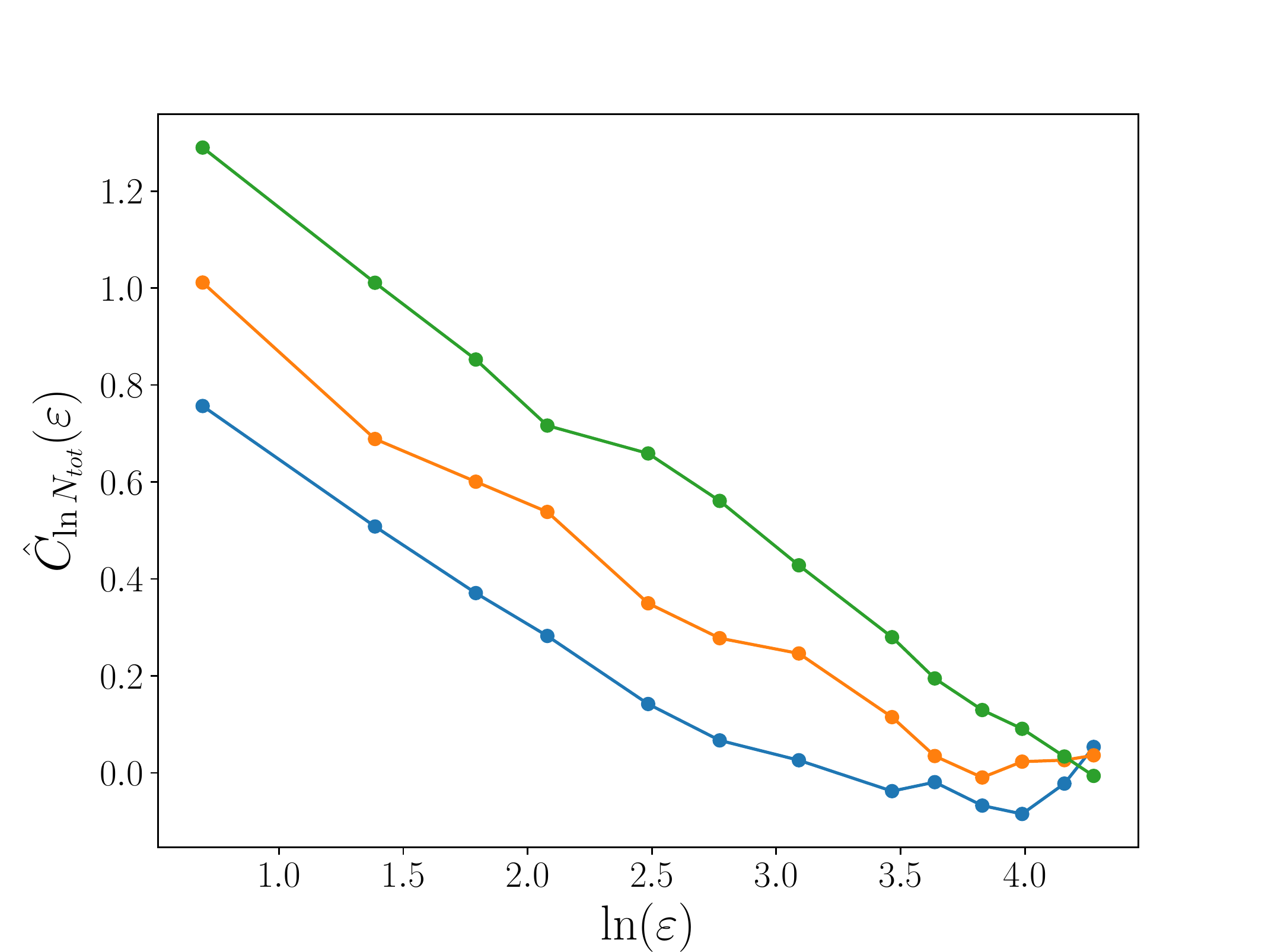}
		\caption{Empirical covariance (Eq. \eqref{eq:emp_cov}) of the logarithmic intensity surrogate as a function of the logarithm of the lag $\varepsilon$ for Corsica (blue), LR (orange) and PACA (green) regions. A multiplicative bias correction calibrated using a synthetic cascade model as been applied to each curve.}
		\label{fig:covlog}
	\end{figure}
\end{center}

\vspace*{-1cm}
As emphasized in (\cite{mrw1,mrw2}), a simple method to estimate the integral scale and the intermittency coefficient is to study the spatial dependence covariance of $\ln \Lambda(B_\ell(\vx))$ that,
according to the multifractal cascade picture, should behave  when $\ell \leq |\vx_1-\vx_2| \leq L$ as:
\begin{equation}
\label{eq:covlog}
  Cov \left\{ \ln \Lambda(B_\ell(\vx_1)) , \ln \Lambda(B_\ell(\vx_2))   \right\} \simeq \lambda^2 \ln \frac{L}{|\vx_1-\vx_2|}
\end{equation}
where $B_\ell$ is a small box of size $\ell$ and the covariance has to be understood as computed under the law of spatial fluctuations.
This equation, which can be shown to be a direct consequence of the existence of a multiplicative cascade process (\cite{mrw1,mrw2}), means that when one plots the covariance of the logarithm of the intensity as a function of the logarithm of the spatial distance,
one gets a straight line of slope $-\lambda^2$ and intercept $\ln L$ (see also \cite{baimu2010}).
Since the values of intensity field $d \Lambda(\vx)$ are not observable (which precisely motivated the use of EM method for the moment estimation) one cannot directly check the validity of Eq. \eqref{eq:covlog}. However, we can use the surrogate intensity $\tL_k$ introduced in Sec. \ref{sec:m1} (Eq. \eqref{eq:proxy})
by collecting all the ignition events at a given spatial location over the whole
period of 27 years and estimate expression
\eqref{eq:covlog} as:
\begin{equation}
\label{eq:emp_cov}
 {\widehat C}_{\ln N_{tot}}(\varepsilon) = \left\langle \ln \tL(B_2(\vx_1)) \ln \tL(B_2(\vx_2))  \biggr \rvert |\vx_1 - \vx_2| = \varepsilon \right\rangle -\left\langle \ln \tL(B_2(\vx))  \right\rangle^2
\end{equation}
where again $\langle . \rangle$ has to be understood as a mean as respect to spatial positions over
the support of $\Lambda$.
It is noteworthy that, when computing such an empirical covariance, one observes a bias
as respect to the true intermittency coefficient that depends on the amplitude of $\langle \Lambda_k \rangle$. Indeed, the greatest the intensity, the smallest the size of relative fluctuations of $\tL_k$
and therefore the smallest this bias. The exact dependence of this bias as a function of $\langle \Lambda_k \rangle$ can be hardly expressed analytically and has been calibrated using the toy model described in Sec. \ref{ssec:mll_ex}.
In Fig. \ref{fig:covlog}, we have plotted the bias-corrected empirical covariance \eqref{eq:emp_cov} as obtained from the three regions data. It is striking that in for all 3 regions, the logarithm
of the local intensities appears to be strongly spatially correlated over large distances
with a correlation function that decreases logarithmically, precisely as one expects for a random cascade model (Eq. \eqref{eq:covlog}).
The value $\lambda^2 \simeq 0.4$ estimated from the intensity covariance logarithmic decay is consistent with former value obtained from the moment scaling properties.
This logarithmic dependence also allows us to measure an approximate value of the integral scales $L$ as
the intercept of the empirical curves. We find respectively $L_{Corsica} \simeq 30$ km,
$L_{LR} \simeq 50$ km and $L_{PACA} \simeq 90$ km. Notice however that the uncertainty of these values is quite large due to the number of possible biases (linked to the assumptions we made, the quality of the data,....) and to the finite sample statistical fluctuations. We can say that the order of magnitude of the integral scale is around $50$ km.

\begin{center}
	\begin{figure}[t]
		\includegraphics[width=0.97 \textwidth]{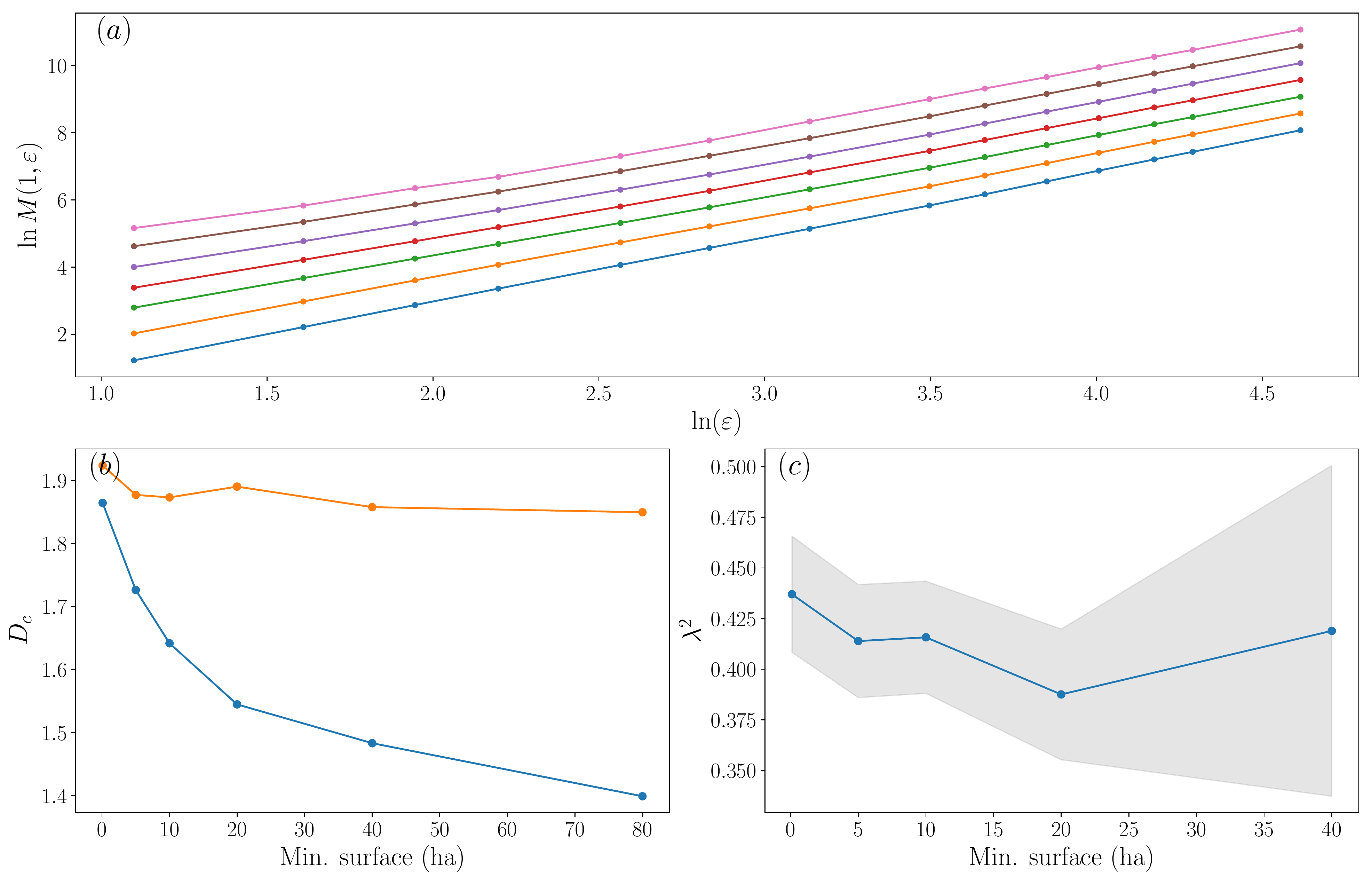}
		\caption{Estimation of fractal dimension and the intermittency coefficient as a function of the wildfire sizes. (a) $\ln M(1,\varepsilon)$ as a function of $\ln \varepsilon$ for all burnt area thresholds from 0.1 Ha (blue bottom curve) to 80 Ha (pink top curve). Curves have been arbitrary shifted for the sake of presentation. (b) Estimated $D_c$ as a function of the area threshold. $D_c$ is estimated at both fine scales (blue curve) and large scales (orange curve). (c) Estimated intermittency coefficient $\lambda^2$ as a function of the burnt surface threshold. The shaded region indicates the order of magnitude of the estimation variations when one changes the fitting domain from small to large scales. }
		\label{fig:Spectra_vs_Size}
	\end{figure}
\end{center}

\vspace*{-1cm}
To conclude our empirical study of multifractal properties of fire ignition spatial distribution,
we have checked that the spectra $\zeta_q$ and $f(\alpha)$ are stable over
the considered time period by performing estimations over a sliding windows of 6 years
from 1992 to 2012. Up to some fluctuations for the highest and lowest singularities (i.e., related to tail events) we observed that the results are consistent. We also checked how the observed scaling properties may depend on the event intensities, namely on the size of the considered wildfires.
We have reproduced the same multifractal analysis for fire events in Corsica but by selecting events with burnt area greater than 
respectively 1, 5, 10, 20, 40 and 80 ha. Since for large areas, the number 
of corresponding events is quite small (only 200 events for fires greater than 80 ha and 320 events for those greater than 40 ha over the whole period 1992-2018), the full $\zeta_q$ spectrum can be hardly estimated. For that reason, we only reported in Fig. \ref{fig:Spectra_vs_Size} the values of the estimated dimension of the support and the values of the intermittency coefficient.
In Fig. \ref{fig:Spectra_vs_Size}(a), we have plotted $\ln M(1,\varepsilon)$
as a function of $\ln(\varepsilon)$ as obtained for various surface thresholds. The slopes of these curves provide a direct estimation of the fractal dimension of the support of intensity (see Eq. \eqref{eq:scaling_M1}). We see that, whereas all curves are parallel at large scales, there is a noticeable threshold dependence at small scales. The values of $D_c$ obtained from large (orange) 
and small (blue) scale fitting are reported as a function of minimum burnt area in Fig. \ref{fig:Spectra_vs_Size}(b). It appears that small scale estimation confirms the observations of \cite{TELESCA20071326} who noticed a strong dependence of the dimension as respect to the event sizes. However, at larger scales the fractal dimensions appear to no longer depend on the threshold. These observations (that agree with the findings of \cite{KANEVSKI2017400}) mean that either the set of large fire locations is fractal only at small scales or that the smaller the number of observed events of a fractal Poisson process, the more biased is the estimation of its fractal dimension at small scales. A specific study devoted to the issue of estimating the fractal dimension of a low intensity point process will be carried out in a future work. In Fig. \ref{fig:Spectra_vs_Size}(c), we plotted the estimated intermittency coefficients for each burnt area threshold. 
Despite the large uncertainty in the estimates, our observations suggest that, whatever their sizes, the fire occurrence is a multifractal point process with an intermittency coefficient close to 0.4.
    
\section{Summary and prospects}
In this paper we presented a new method to estimate the multifractal properties
of point patterns with clustering features when these latter result from the spatial fluctuations
of the expectation measure and not from peculiar correlations in the event occurrence likelihood.
The paradigm of such process is a spatial Cox processes with an intensity measure that is provided by a random
cascade model. When only a few number of realizations are available so that
the intensity measure remains unknown, we have shown that the moments of this measure at each scale can be still be estimated through a maximum likelihood approach that consists of representing the observed distribution number of events as a mixture of simpler
distributions (like e.g. Poisson-Log-Normal or Negative Binomial distributions). The model calibration can then be performed using  a classical Expectation Maximization procedure. Our approach has been validated on mono- and multi-fractal toy models involving an intensity lying on a statistically self-similar and shift-invariant random Cantor sets.

We have applied this framework to the annual wildfire ignition events of French Mediterranean regions
gathered in the Prom\'eth\'ee database. 
Our study suggests that the clustering features of the wildfire distributions do not result from peculiar correlations in the event occurrence likelihood but reflect the multifractal spatial structure of the intensity.
Indeed, we have shown that the 
inhomogeneous Ripley $L(r)$ function behavior is consistent with 
a ``Complete Spatial Randomness'' situation. Moreover, 
the $\zeta_q$ exponent spectrum governing the power-law behavior
of the order $q$ moment of the intensity distribution
behaves as a strictly concave function, the hallmark of multifractal processes.
All the three studied regions exhibit almost the same multifractal features:
A dimension of the support close to $D_c \simeq 1.9$ and a singularity spectrum extending
from $\alpha \simeq 1.2$, for the locations
with highest intensities, to  $\alpha \simeq 3$ for the locations with weakest intensities, the most probable values being around $\alpha = 2$.
In all regions, the intermittency coefficients have almost the same values
$\lambda^2 \simeq 0.4$ with a large correlation scale around $50$ km.
Moreover, it appears that these multifractal features do not depend
on the time period chosen nor on the size of selected fire events.

These results show that the spatial distribution of fire ignition is of complex nature
and quantifying or comparing the fire ignition hazard is not a trivial task.
In a future work, we plan to consider possible applications of our approach to provide practical help-to-decision tools for the prevention of wildfires, to monitor various kinds of firefighting policies or forest management strategies and to quantify extreme events in relationship with global warming effects.
For instance, one could exploit the scaling pre-factors values (the constant factors in scaling expressions) in order to design robust metrics for
comparing ignition risks between different regions. One could also use the parametric description of intensity laws at various scales in order to estimate the likelihood of occurrence of extreme values.
Finally, from a fundamental point of view, it remains to understand and interpret our findings. Since a multifractal random cascade field $X$ is basically built as a product $X = \prod_{i=1}^N W_i$ where
fields $W_i$ are independent random processes correlated over a scale $r_i = C r^i$, one could wonder if the intensity associated with a fire ignition event in a given small cell could not be written in such a way. Indeed, because the intensity measure of a small cell is roughly the probability to observe one event in this cell ($\Lambda(dx) \simeq Prob \{N(dx)=1\}$), it is tempting to decompose this probability
as the product of a large number of probabilities associated
with the various independent factors that may impact the ignition likelihood
(for instance the nature of the vegetation, the accessibility of the considered site, the number of visits of this site, the meteorological factors,...). These factors may display a wide spectrum of correlation lengths or even may be themselves self-similar which would explain the scaling properties we have estimated.
An empirical analysis of the multifractal properties of ignition intensity fields observed in other regions worldwide where such factors may strongly vary could help to obtain a better understanding of our results.

Finally, let us underline that many statistical aspects related to the estimation of  multifractal point processes remain to be investigated. For example, as illustrated by our discussion in \ref{App1}, the simple question of the estimation of the fractal dimension of the set that supports the point process, notably when its intensity is weak, deserves to be explored more deeply.

\section*{Acknowledgements}
The authors are grateful to all anonymous referees for their constructive remarks and comments on the first version of the paper.

\appendix

\section{Estimating the fractal dimension of the intensity measure support $\cS$ using correlation integral and sandbox methods.}
\label{App1}
In many studies involving spatial point process distributed on a fractal structure of dimension $D_c$, the authors considered either the correlation function or the sandbox method in order to estimate $D_c$. Let us show, using an heuristic argument, that, when the point process
is multifractal, these methods provide a biased estimation.
The correlation integral, introduced to study chaotic systems and notably used by \cite{TELESCA20071326}, consists in measuring the scaling behavior of $C(\varepsilon)$ that counts 
the number of pair of events separated by a distance smaller than $\varepsilon$. The sandbox 
method used for example in \cite{KANEVSKI2017400}, consists in estimating $D_c$ from the scaling
of $M(\varepsilon)$, the average number of neighbor events at a distance smaller than $\varepsilon$ 
from a given event.

Let us consider a multifractal Cox process with a multifractal intensity $d \Lambda(x)$ as defined in Sec. \ref{sec:mpp}.
To simplify our purpose, we suppose that one observes a total of $N_T$ events and we 
locate them on a grid of mesh $\varepsilon$. Let $B_\varepsilon(k)$ the grid cells where $k=1, \ldots, N_\varepsilon$. If $i$ is the $i$-th observed event, we denote by $B_\varepsilon[k(i)]$ the box of the grid of index $k$ that contains event $i$.

Then one has, at resolution $\varepsilon$: 
$$
 M(\varepsilon) \simeq N_T^{-1} \sum_{i=1}^{N_T} N\left[B_\varepsilon(k(i))\right] 
$$
By re-indexing the sum over the grid cell numbers, one gets:
\begin{eqnarray*}
M(\varepsilon) & \simeq & N_T^{-1} \sum_{k=1}^{N_\varepsilon} \sum_{i=1}^{N\left[B_\varepsilon(k)\right]} N\left[B_\varepsilon(k)\right] \\
& = &  N_T^{-1} \sum_{k=1}^{N_\varepsilon} N^2\left[B_\varepsilon(k)\right]
\end{eqnarray*}

Along the same line, one has:
\begin{eqnarray*}
	C(\varepsilon) & \simeq & \frac{1}{2} N_T^{-1} \sum_{i=1}^{N_T} (N\left[B_\varepsilon(k(i))\right]-1) \\
	 & = & \frac{1}{2}  N_T^{-1} \sum_{k=1}^{N_\varepsilon} \sum_{i=1}^{N\left[B_\varepsilon(k)\right]} (N\left[B_\varepsilon(k)\right]-1) \\
	& = &  \frac{1}{2}  N_T^{-1} \sum_{k=1}^{N_\varepsilon} N\left[B_\varepsilon(k)\right](N\left[B_\varepsilon(k)\right]-1)
\end{eqnarray*}

If one takes the average of both quantities as respect to the Poisson law, one gets, from the definition 
\eqref{eq:def_zq} of the partition functions $Z(q,\varepsilon)$:
\begin{eqnarray*}
 C(\varepsilon) & \simeq & \sum_{i \in {\cal P}_\varepsilon} \Lambda^2(B_\varepsilon(i)) = Z(2,\varepsilon ) \\
 M(\varepsilon) & \simeq  & \sum_{i \in {\cal P}_\varepsilon} \left(\Lambda^2(B_\varepsilon(i))+\Lambda(B_\varepsilon(i)) \right) = Z(2,\varepsilon)+Z(1,\varepsilon)
\end{eqnarray*}

which leads to, given that $\tau_1 = 0$:
\begin{eqnarray*}
	C(\varepsilon) & \sim & C \varepsilon^{\tau_2}\\
	M(\varepsilon) &  \sim & C_1 + C_2 \varepsilon^{\tau_2} . 
\end{eqnarray*} 

We see that (i) the behavior at small scales of $M(\varepsilon)$ is not a pure power law because 
it is not a factorial moment and (ii) both $C(\varepsilon)$ and $M(\varepsilon)$ provide biased 
estimate of the fractal dimension if $\Lambda$ is a multifractal measure.
Indeed, in that case $\tau_q$ is strictly concave and therefore $\tau_2+\tau_0 - 2 \tau_1 < 0$. 
From $\tau_1 = 0$ and $\tau_0 = -D_c$, one gets $\tau_2 < D_c$.
For example, in the case of a log-normal cascade of intermittency coefficient $\lambda^2$,
one has 
\begin{equation} 
 \tau_2 = D_c - \lambda^2
\end{equation}

\section{Bias of standard multifractal methods for low intensity point processes}
\label{App2}

\begin{center}
	\begin{figure}[t]
		\includegraphics[width=0.97 \textwidth]{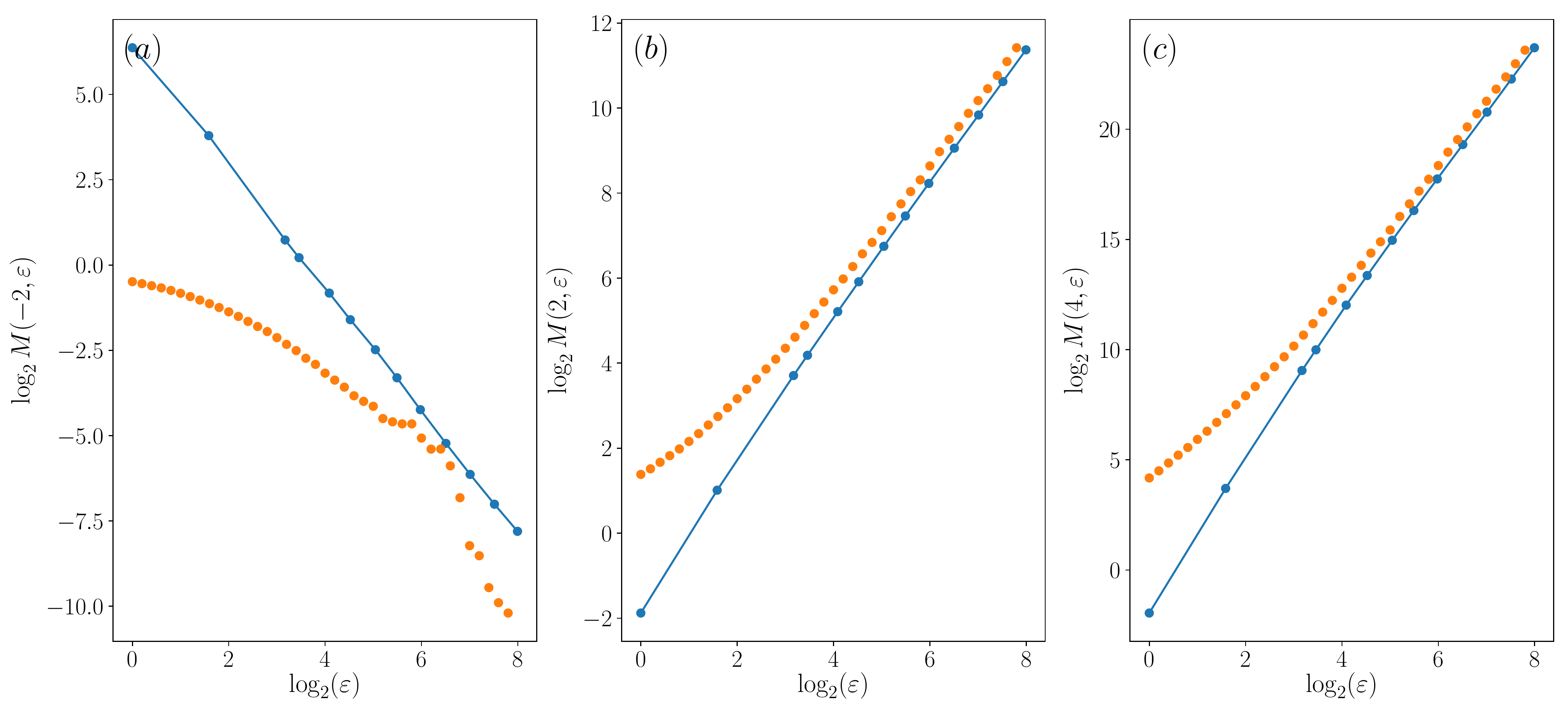}
		\caption{Bias in the WTMM method moment scaling for low intensity Cox process. For $q=-2$ (a), $q=2$ (b) and $q=4$ (c), the scaling behavior of order $q$ WTMM moments (orange points) are compared to the ones defined in Eq. \eqref{eq:def_mq}
		estimated by our EM method (blue points) for the example displayed in Fig  \ref{fig:ex1}. One sees that WTMM is strongly biased for negative or small positive $q$ values.}
		\label{fig:wtmm_vs_fmoments}
	\end{figure}
\end{center}

If $N$ follows a Poisson law of intensity $\Lambda$, it is well known
that the factorial moments for any $q \in \mathbb{N}$ satisfy:
$$
 E(N^{[q]})  = \Lambda^q
$$
where $N^{[q]} = N(N-1) \ldots (N-q+1)$. If
$\Lambda$ is stochastic, we then have:
$$
\langle N^{[q]} \rangle  = \langle \Lambda^q \rangle \; .
$$
It results that the generalized moments of $\Lambda$ can be interpreted as factorial moments of $N$ with a continuous order.

Moreover, since,
$$ N^{[n]} = \sum_{k=1}^n s_{n,k} N^k$$
where $s_{n,k}$ are constants (the Stirling numbers),
one will have:
$$\langle N(B_\varepsilon(\vx_i))^n \rangle  = \sum_{k=1}^n s_{n,k} \langle \Lambda (B_\varepsilon(\vx_i))^k \rangle  \; .$$
For a multifractal Cox process such that 
$$\langle \Lambda(B_\varepsilon(\vx_i))^n \rangle \simeq C_n \varepsilon^{\zeta_n}, $$
it will result that  $\langle N(B_\varepsilon(\vx_i))^n \rangle$ does not satisfy an exact 
scaling but is a mixture of all power-laws ranging from  $\langle \Lambda(B_\varepsilon(\vx_i)) \rangle \sim C_1 \varepsilon^{\zeta_1}$ to $\langle \Lambda(B_\varepsilon(\vx_i))^n \rangle \sim C_n \varepsilon^{\zeta_n}$. When $\Lambda < 1$ the first term dominates while when $\Lambda > 1$ the last term is the largest. If $\Lambda(B_\varepsilon(\vx_i))$ is small enough at small scales, since it increases when $\varepsilon$ increases, 
one expects a cross-over from $\varepsilon^{\zeta_1}$ at fine scales to $\varepsilon^{\zeta_n}$ at 
coarse scales. Such cross-over behavior can be expected for any method that will rely on the scaling 
properties of the observable events $dN(\vx)$.
This is illustrated in Fig. \ref{fig:wtmm_vs_fmoments} on the 1D log-normal example of section \ref{ssec:mll_ex} displayed in Fig. \ref{fig:ex1}. We have reported a comparison 
between the scaling properties of $M(q,\varepsilon)$ computed from the estimated law of 
$\Lambda$ as explained in Sec. \ref{ssec:mll} (Eq. \eqref{eq:def_mq}) and the partition function
of the WTMM method (see \cite{wtmm1,wtmm2}), applied to the sample of $N(dx)$.
We have displayed in log-log representation $M(q,\varepsilon)$ (blue curves) and $ Z_{wtmm}(q,\varepsilon)/Z_{wtmm}(0,\varepsilon)$ (orange curves) for $q=-2,2,4$. Curves have been shifted by arbitrary constants for clarity purpose. One clearly
sees that at small scales, the scaling exponents of $Z(q,\varepsilon)$ is very small, while at larger 
scales both methods tend to provide the same estimations. It is noteworthy that the larger the $q$ values the larger the domain of scales where the WTMM partition functions are not biased. This can be explained by the fact that for large $q$, the largest values of $\Lambda(dx)$ are involved and therefore the difference between moments of $N$ and $\Lambda$ become negligible at intermediate scales. On the other hand, for negative $q$, the smallest values of $\Lambda(dx)$ are governing 
the partition function behavior and even at large scales, the WTMM estimates are not reliable.

\section{Clustering properties and the Poisson hypothesis}
\label{clusterProp}

\begin{center}
	\begin{figure}[t]
		\hspace*{2cm}
		\includegraphics[width=10cm]{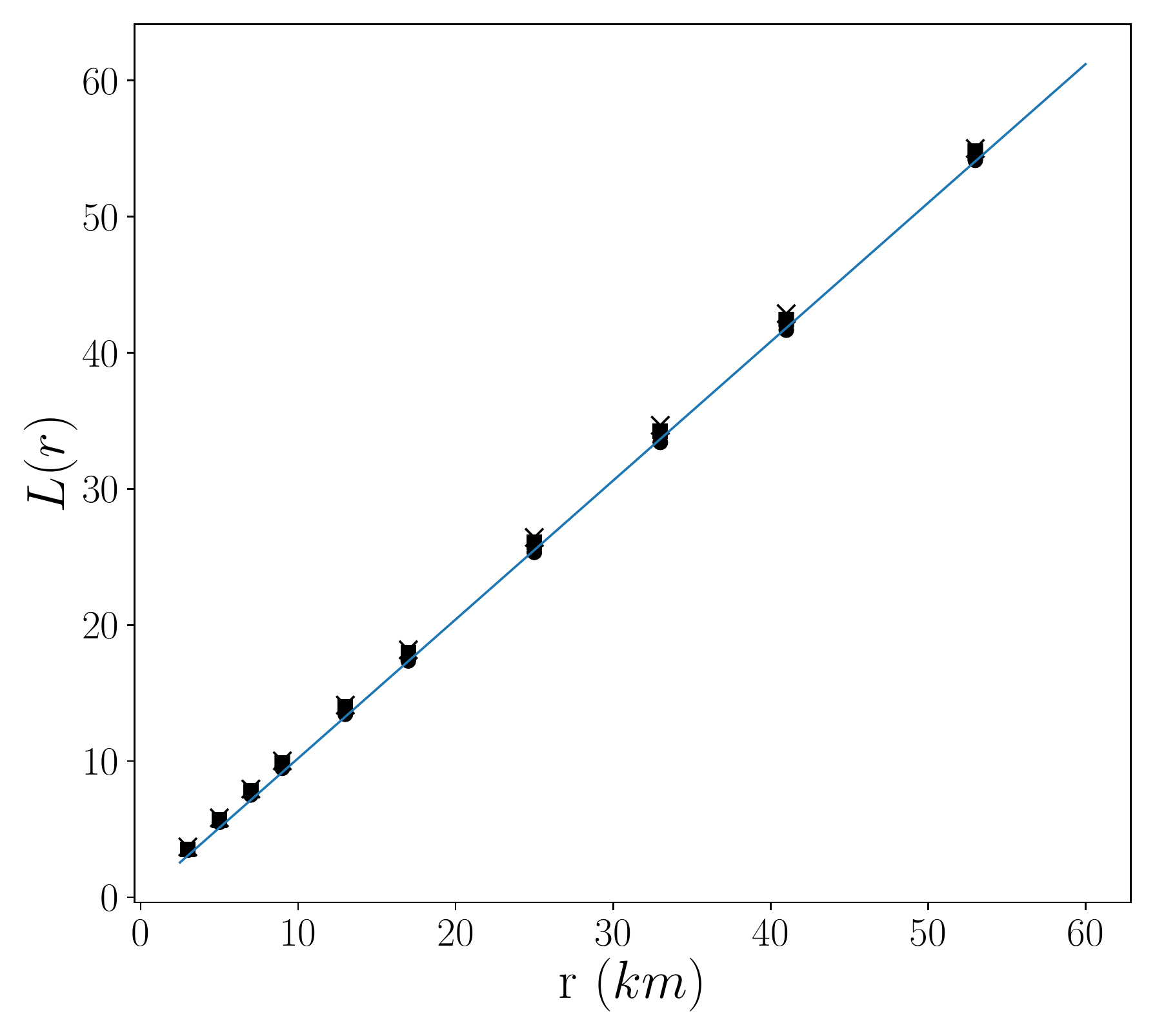}
		\caption{Estimated Ripley inhomogeneous $L(r)=\sqrt{K(r)}$ function for
			Corsica,PACA and LR regions. The solid line stands for $L(r)=r$ as expected
			for uncorrelated event numbers $N_k$'s.}
		\label{fig:Ripley}
		
	\end{figure}
\end{center}

In order to describe the spatial fluctuations of the annual
number of fire occurrence $dN(x,y)$, we will  use the approach described in Sec. \ref{sec:mpp}, where we suppose that $dN(x,y)$ is a Cox process, i.e., conditionally to random multifractal spatial intensity $d\Lambda(x,y)$, $dN(x,y)$ is an inhomogeneous Poisson process.
This assumption notably implies that, for any given the intensity
function, the observed number of ignitions during a given year over distinct areas are uncorrelated.
This means that the observed
spatial clustering of ignition locations is exclusively due to the spatial fluctuations of the intensity field. In order to check for such a feature we follow the method proposed in \cite{Hering2009} where the authors
define a inhomogeneous version of the Ripley $K$-function that allows one to filter out the spatial dependence of the intensity and test to remaining existing correlations.
Accordingly, one defines:
\begin{equation}
\label{ripley}
K(r) =  S^{-1} \sum_{\vx_k \in {\cal S}} \; \; \sum_{\vx_j \in B_r(\vx_k) \cap {\cal S}\backslash\vx_k} \frac{ E \left[ N_k N_j \right]}{w_{jk} \tL_k \tL_j}
\end{equation}
where $\cal S$, $S$, $N_k$, $\tL_k$ and the expectation $E(.)$ are defined in Sec. \ref{sec:m1} and $w_{kj}$ is a Ripley edge correction factor designed to correct biases caused by the edges of
the domain (see \cite{Hering2009})\footnote{Actually, we adapted the usual Ripley correction factor from circle geometry to square geometry. If $d_{kj} = \max(|x_j-x_k|,|y_j-y_k|)$ stands for the square distance, $w_{kj}$ represents the fraction of DFCI squares $m$ at distance $d_{kj}$ from $\vx_k$ that are in the studied
	region.}.
Notice that in absence of correlation, $E \left[ N_k N_j \right] = \tL_k \tL_j$
and therefore $L(r) = \sqrt{K(r)} = r$.
We have computed $L(r)$ according to expression \eqref{ripley}
for the 3 regions. As it can be seen in Fig. \ref{fig:Ripley}, the plots of $L(r)$ closely follow the straight line
$L(r)=r$ in the 3 cases which suggests that the random variables $N_k$ and $N_j$
are uncorrelated. This result confirms the finding of \cite{Hering2009} and shows that observations are compatible with a Cox process. Notice that it might be quite surprising that one does not observe any 
spatial anti-correlation between fire occurrence events since one expects that after a wildfire another 
one cannot occur nearby within an already burnt area. However one has to remind that the minimal considered surface in our study is $4 km^2$ which is quite huge as respect to the typical wildfire burnt areas.
Very large fires that will contribute to an effective anti-correlation are very few and statistically insignificant.

\section*{References}










\vskip 1 cm

\end{document}